\documentclass[aps,prd,epsf,showpacs,amsmath,amssymb,graphics,floatfix,onecolumn,11pt]{revtex4}
\usepackage{graphicx}
\usepackage{dcolumn}
\usepackage{bm}
\usepackage{color}
\usepackage{amsmath}
\usepackage{array}

\newcommand{\be}{\begin{equation}}
\newcommand{\ee}{\end{equation}}
\newcommand{\ba}{\begin{eqnarray}}
\newcommand{\ea}{\end{eqnarray}}
\newcommand{\ban}{\begin{eqnarray*}}
\newcommand{\ean}{\end{eqnarray*}}

\newcommand{\eq}[1]{(\ref{#1})}

\begin{document}
\title{Infinite efficiency of collisional Penrose process: 
Can over-spinning Kerr geometry be the source of ultra-high-energy cosmic rays and neutrinos ?}

\author{$^1$Mandar Patil\footnote{Electronic address: mandar@rikkyo.ac.jp, mandarppatil@gmail.com},
$^1$Tomohiro Harada\footnote{Electronic address: harada@rikkyo.ac.jp}, \\
$^2$Ken-ichi Nakao\footnote{Electronic address: knakao@sci.osaka-cu.ac.jp},
$^3$Pankaj S. Joshi\footnote{Electronic address: psj@tifr.res.in},
$^{4,5}$Masashi Kimura\footnote{Electronic address: masashi.kimura@tecnico.ulisboa.pt}
}

\affiliation{  $^{1}$Department of Physics, Rikkyo University,
Toshima-ku, Tokyo 171-8501 Japan. \\ 
$^2$Department of Mathematics and Physics, Graduate School of Science, 
Osaka City University, Osaka 558-8585, Japan. \\ 
$^3$Tata Institute of Fundamental Research, Homi Bhabha Road, Mumbai 400005, India. \\ 
$^4$DAMTP, Centre for Mathematical Sciences, University of Cambridge, Wilberforce Road, Cambridge CB3 0WA, United Kingdom. \\
$^5$CENTRA, Departamento de F\'{\i}sica, Instituto Superior T\'ecnico--IST, 
Universidade de Lisboa--UL, Avenida~Rovisco Pais 1, 1049 Lisboa, Portugal
}


\maketitle

\pagebreak
\section*{\Large{Abstract}}
The origin of the ultra-high-energy particles we receive on the Earth from the outer space such as EeV cosmic 
rays and PeV neutrinos remains an enigma. All mechanisms known to us currently make use of
electromagnetic interaction to accelerate charged particles. In this paper we propose a mechanism 
exclusively based on gravity rather than electromagnetic interaction. We show that it is possible
to generate ultra-high-energy particles starting from
particles with moderate energies using the collisional Penrose process in an 
overspinning Kerr spacetime transcending the Kerr bound only by an infinitesimal 
amount, i.e., with the Kerr parameter $a=M(1+\epsilon)$, where
we take the limit $\epsilon \rightarrow 0^+$. 
We consider two 
massive particles starting from rest at infinity that collide at $r=M$ with divergent center-of-mass energy and produce
two massless particles. 
We show that massless particles produced in the collision can escape to
 infinity with the ultra-high energies exploiting 
the collisional Penrose process with the divergent efficiency
 $\eta \sim {1}/{\sqrt{\epsilon}} \rightarrow \infty$.
Assuming the isotropic emission of massless particles in the center-of-mass frame of the colliding particles, 
we show that half of the particles created in the collisions
escape to infinity with the divergent energies, while the proportion of 
particles that reach infinity with finite energy is minuscule.
To a distant observer, ultra-high-energy particles appear to originate 
from a bright spot which is at the angular location 
$ \xi \sim {2M}/{r_{obs}}$ with respect to the singularity on the side which is rotating towards the observer. 
We compute the spectrum of the high energy massless particles and show that anisotropy in the
emission in the center-of-mass frame leaves a distinct signature on its shape. 
Since the anisotropy is dictated by the 
differential cross section of the underlying particle physics process, the observation of the spectrum can
constrain the particle physics model and serve as a unique probe into fundamental physics at ultra-high energies
at which particles collide. 
Thus, the existence of the near-extremal overspinning Kerr geometry in the Universe, either as a transient 
or permanent configuration, would have deep 
implications on astrophysics as well as fundamental particle physics.

\pagebreak

\section{Introduction}

The Earth is bombarded with cosmic rays from the outer space consisting of protons as well
as nuclei such as iron with energies 
that extend all the way upto several hundreds of EeV \cite{Lisley},\cite{Watson}. 
There is no general consensus as of now on the acceleration 
mechanism that is responsible for  the generation of the EeV particles. Two leading mechanisms are
1. diffusive acceleration by shock waves,
i.e., Fermi acceleration and 2. acceleration by electric fields generated by time varying magnetic fields, 
i.e., unipolar inductors (see \cite{Hillas},\cite{Kotera},\cite{Stanev} and references therein). 
The conditions conducive to the particle acceleration
can be found in a wide variety of astrophysical settings such as neutron stars, gamma ray bursts, active galactic nuclei, 
colliding galaxies and so on. We note that all the proposed acceleration mechanisms make use of electromagnetic
interaction to accelerate charged particles. Cosmic rays with energies 
beyond 60 EeV are expected to produce neutrinos as they interact with
the cosmic microwave background photons \cite{Greisen},\cite{Zatsepin}. Neutrinos with 
the energies in the range 100 TeV to PeV have been observed recently \cite{Aartsen},\cite{Aartsen2}. 
However the connection between the production of ultra-high-energy neutrinos and cosmic
rays is far from being understood \cite{IceCube}. 

The origin of the ultra-high-energy particles continues to be an enigma and a
topic of an active ongoing investigation. Hence it is worthwhile to look for the other mechanisms
that could accelerate particles to high energies. In this paper 
we present a novel mechanism to generate high-energy particles that exclusively makes use 
of gravity rather than electromagnetic interaction. 

The Kerr metric is one of the simplest and most elegant solution to the Einstein equations in general relativity which
exhibits many remarkable features \cite{Kerr}. 
The Kerr solution represents a rotating black hole if the spin parameter is smaller than the mass, i.e., $a \le M$. 
For large enough values
of the spin parameter, i.e., $a>M$, the Kerr metric represents a rotating naked singularity. One of the interesting features 
associated with the Kerr solution is the dragging of inertial frames,
as a consequence of which spacetime can host the particle orbits with negative energies in the region called ergosphere. 
It was argued by Penrose that if a particle disintegrates into two 
particles, wherein one of the particles is launched onto the orbit with negative energy,
the other particle can have energy larger than 
that of the initial particle. This process is known as Penrose process \cite{Penrose1},\cite{Penrose2}. 
Thus, one can throw in a particle into the ergosphere from the distant location and 
get out a particle with larger energy. An analogous process was considered 
wherein two particles collide and scatter. One of the scattered particles moves along the negative energy orbit,
whereas the other particle is now endowed with energy larger than the combined energy 
of the two colliding particles. This process is known as
the collisional Penrose process. It was argued that the efficiency of the Penrose process and collisional
Penrose process, defined as the ratio of the energy output to the energy input admits an upper bound
which is around 1.2 for Kerr black hole \cite{Bardeen},\cite{Wald},\cite{Piran1},\cite{Piran2}.

Interest in the collisional Penrose process was revived recently in light of the re-discovery of 
the process of ultra-high-energy particle collision around the extremal Kerr black holes \cite{BSW},\cite{Piran2},\cite{Harada}.
When a particle with a specific critical value of the
angular momentum, which asymptotically
approaches the event horizon of the extremal Kerr black hole, collides at the location close to event horizon, with
another particle with a sub-critical angular momentum that
is radially ingoing at the horizon, the center-of-mass energy of collision shows divergence.
This is relevant from the point of view of particle physics processes such as dark matter annihilation 
for which the cross section is dismal at low energies and is expected to show an upward trend at
high energies or the resonances which occur at high energies \cite{Cannoni},\cite{Arina}. Debris from the 
ultra-high-energy collisions around the extremal Kerr black hole escaping to infinity is expected to have imprinted on
it the signature of the physics at high energies. It was shown that the energies of the particles which are generated 
in the ultra-high-energy collisions around the Kerr black hole that manage to escape to infinity would be finite
despite the boost by the collisional Penrose process \cite{Bejger},\cite{Harada1}. The efficiency of 
the collisional Penrose process can go upto 1.4. The escape fraction, i.e., the fraction of the particles produced in 
the ultra-high-energy collision that escape to infinity was shown to be vanishingly 
small \cite{BSW2},\cite{Williams},\cite{McWiliams}. Thus, the significance 
of the process of ultra-high-energy collisions around Kerr black holes from an observational point of view is not very clear.

Recently, it was shown that the efficiency of the collisional Penrose process can be boosted up by one order of magnitude by 
considering a collision between the particle with angular momentum slightly larger than the critical value, 
which turns back as an outgoing particle with the vanishingly small outward velocity just 
outside the horizon due to the angular momentum barrier and a sub-critical ingoing particle.
The critical particle was taken to be an ingoing particle at the collision
in the earlier investigations. The upper bound on the efficiency was revised to be around 14 \cite{Schnittman}.

The collision between an outgoing sub-critical particle and ingoing sub-critical particle just outside the horizon of the
extremal Kerr black hole was considered in \cite{Berti}. The efficiency of the collisional Penrose process was shown to
diverge in this case. However, the outgoing sub-critical particle must be produced in yet another 
collision just outside the horizon, 
since such a particle cannot start from the distant location and then turn back as an outgoing 
particle just outside the horizon and also it cannot emerge from the 
black hole as nothing comes out of the black hole. It was shown that it would be possible to produce an outgoing 
sub-critical particle in the preceding collision only if one of the colliding particles is super-heavy 
with divergent mass. Thus, the overall efficiency in the process of multiple collisions turns out to be finite and
can take values upto 14. We, however note that the efficiency of the collisional Penrose process in 
the extremal Reissner-Nordstr\"{o}m black hole spacetime case can be arbitrarily 
large\cite{Zaslavskii1},\cite{Zaslavskii2},\cite{Nemoto}.

Thus, it is very unlikely that it would be possible to extract a large amount of energy from the Kerr black hole. 
In this paper, we go beyond the extremality and consider the over-spinning Kerr geometry. 
The spacetime does not admit an event horizon but has a singularity at $r=0$ in the equatorial plane,
which is visible to a distant observer. There is a possibility that an overspinning Kerr geometry
could arise as a transient configuration in the region exterior to the 
regular matter cloud that undergoes a gravitational collapse followed by a bounce, thereby avoiding the occurrence of
the singularity \cite{Nakao}. It might also be possible to overspin a near-extremal Kerr black hole with 
a test particle \cite{Jacobson}. However the self-force could act as a cosmic censor in 
this process \cite{Barausse},\cite{Colleoni}.
Recently it was argued that there are a number of ways by which the overspinning spacetime geometries can arise 
in the context of string theory which provides resolution of naked singularities \cite{Gimon}. 
For instance, $5+1$ dimensional solutions to the heterotic string theory 
presented in \cite{Horowitz}, when looked at from the large distance, appear to be $3+1$ dimensional spacetimes
sourced by the overspinning naked singularities. When the singularity is approached, the two compactified extra 
spatial dimensions manifest themselves and the $3+1$ 
dimensional singularity is resolved. The over-spinning naked singular geometry has been shown to be unstable to 
the linear perturbations \cite{Dotti},\cite{Cardoso}. 
Their analysis assumes the spacetime to be $3+1$ dimensional and that the overspinning Kerr geometry extends
either throughout the spacetime or is valid outside the hyper-spinning object with the ingoing or reflecting 
boundary conditions imposed on its surface.  We note that the imaginary part of the angular frequency $\omega_I$ of the gravitational wave perturbation 
approaches zero from positive value in the extremal limit \cite{Pani}. 
This means that the near-extremal over-spinning Kerr geometry, which we deal with in this paper, turns out to be
effectively marginally stable.
Further it remains to be investigated whether the overspinning geometries arising in the 
context of string theory via non-trivial ways
are unstable to the perturbations. The high energy completion of the Kerr naked singular geometry could stabilize 
it against the perturbations.

In this paper, we consider the collisional Penrose process in the
overspinning Kerr geometry with the spin parameter $a$
transcending the extremal value $M$ by a small amount, i.e., 
$a=M\left(1+\epsilon\right)$, where we take the limit 
$\epsilon \rightarrow 0^{+}$. Here 
$a$ is assumed to be positive without loss of generality.
We show that it is necessary for the center-of-mass energy of the colliding particles to diverge in order for the 
efficiency of the collisional Penrose process to diverge. We had investigated 
the process of ultra-high-energy particle collisions in the over-spinning Kerr geometry, wherein the ingoing and outgoing
particles, both with the finite radial velocity, collide at $r=M$ with the divergent 
center-of-mass energies in the near-extremal limit \cite{Patil1},\cite{Patil2}.
An initially ingoing particle starting from the distant location enters the circle $r=M$ and 
then turns back as an outgoing particle at the lower radial coordinate if its angular momentum is in the appropriate range. 
Thus{, the outgoing particle that participates in the collision arises
naturally in the overspinning Kerr geometry. Many of the drawbacks associated with the process of ultra-high-energy collision
in the Kerr black hole case are circumvented. The finetuning of the geodesic parameters of the colliding 
particles is not required \cite{Patil1}. The time required for the collision is significantly reduced \cite{Patil3}. 
The upper bound on the center-of-mass energy due to the conservative backreaction 
of the colliding particles is significantly higher \cite{Patil4}. 
The escape fraction of the colliding particles is finite \cite{Patil5}.

We consider a process where two colliding particles scatter and produce two massless particles.
All particles are assumed to move along the equatorial plane for simplicity. It allows us to carry out fully analytical calculations. 
We choose to work in the center-of-mass tetrad. Since the cross section of the particle physics processes
is computed in the center-of-mass frame, it makes the analysis of escape 
fraction and spectrum of the massless particle easier \cite{BSW2}.
The colliding particles travel in the opposite directions in the center-of-mass frame and so do the massless particles which 
are produced in the collision. 
The direction along which the massless particles travel can be oriented at any angle with respect to the direction
along which colliding particles travel in the center-of-mass frame. 
However, the angular distribution of massless particles is dictated by the cross section 
of the underlying particle physics process. 
We compute the conserved energy of the massless particle as a function of this angle.
If a particle escapes to infinity, its conserved energy would be the energy as measured by a distant observer. 
By analyzing the geodesic motion, we determine the escape cones within which the massless particles must be 
emitted so that it reaches infinity. The escape cones span half of the angular range.
We show that the particle emitted along almost all the directions within the escape cones 
reaches infinity with divergent energy. The massless particle which 
travels in the opposite direction with respect to the 
ultra-high-energy particle that escapes to infinity, has a conserved energy which is negative and also divergent.
This implies that the collisional Penrose process is at work and its efficiency shows divergence. It 
is possible to create ultra-high-energy particles by extracting a large energy from the over-spinning Kerr geometry. 
Interestingly, the results we obtain here do not depend on whether the center-of-mass frame moves radially inwards, 
radially outwards or has a zero radial velocity, unlike 
in the black hole case\cite{Berti}.

All ultra-high-energy massless particles produced in the high-energy collisions 
have positive angular momenta and almost identical
value of the impact parameter. Thus, they co-rotate with the singularity. 
As seen by the distant observer, they seem to appear from a bright spot located at the 
specific angular location on that side of 
the singularity which is rotating towards the observer.
We compute the spectrum of the massless particles with the assumptions of the uniform distribution for
the angular momenta of the colliding 
particles and the cross section for the scattering process being constant at large center-of-mass energies.
The spectrum admits an upper cut-off energy which can however go to infinity in the near-extremal limit. We consider two cases
where the distribution of the massless particles is isotropic in the center-of-mass frame and the case where it is anisotropic.
We show that the anisotropy leaves a distinct imprint on the spectrum. Since anisotropy is determined by the differential
cross section of the underlying particle physics process, the observation of 
the spectrum will allow us to put constraints on and distinguish 
different particle physics models and serve as a probe into fundamental physics at high energies at which particles collide. 
Thus, the existence of the near-extremal overspinning Kerr spacetime
geometry in the Universe, either as a transient configuration or
permanent configuration, would have deep implications for astrophysics as well as fundamental particle physics. 

When the backreaction is taken into account we expect that the over-spinning Kerr geometry would be driven towards the extremal Kerr black hole
configuration. The calculation we present in this paper is meaningful so long as the final spin parameter is larger than the extremal 
value. This would put an upper bound on the energy of the massless particle. We demonstrate that the upper bound is still so high that it can be of great
interest from astrophysical and particle physics point of view.

Interestingly the possibility of production of particle with large conserved energy was briefly mentioned 
in \cite{Zaslavskii3}. We however note that the production of particle with large conserved energy 
by itself does not imply the large efficiency of collisional Penrose process. 
Additionally one must ensure that the particle with large conserved energy escapes to infinity,
which may not always be the case. One should also demonstrate that the colliding particles must occur 
naturally, e.g., starting from rest at infinity.
This requires the assertion on the global behavior of the metric functions as 
opposed to the local analysis presented in the paper above. In this paper we carry out an in depth analysis 
demonstrating that the colliding particles occur naturally, collision results in the production of the particle 
with large conserved energy and this particle indeed escapes to infinity.

\section{Collisional Penrose Process in General Kerr Geometry}

In this section we describe the method we employ to analyze the collisional Penrose process.
We keep our discussion general and deal with the Kerr spacetime with an arbitrary spin parameter. Later on we specialize to the 
overspinning Kerr geometry. We find it convenient to make a transition to the center-of-mass tetrad since the 
cross section of the particle physics processes is specified in the center-of-mass frame which makes the calculation 
of escape fraction and energy distribution of the particles produced in
the collision easier. Initially, we move over to the 
locally non-rotating frame (LNRF) and then specify a Lorentz transformation that relates LNRF and center-of-mass frame.

\subsection{Kerr metric and geodesics in Kerr spacetime}
The Kerr metric in the Boyer-Lindquist coordinate system $\left(t,r,\theta,\phi\right)$ is given by 
\begin{equation}
ds^2 = -\left(1-\frac{2Mr}{\Sigma}\right)dt^2+\left(r^2+a^2+\frac{2Mra^2}{\Sigma} \sin^2 \theta\right) \sin^2 \theta d\phi^2 
-\frac{2Mra}{\Sigma} \sin^2\theta dtd\phi+ \frac{\Sigma}{\Delta}dr^2+ \Sigma d\theta^2 \ ,
\label{Kerrmetric}
\end{equation}
where $\Delta=\left(r^2-2Mr+a^2\right)$ and $\Sigma=\left(r^2+a^2\cos^2\theta\right)$. The Kerr metric contains two 
parameters, namely mass $M$ and spin parameter $a= {J}/{M}$ where $J$ is the angular momentum. Without the loss
of generality we assume that the spin parameter $a$ is positive. When $a
\le M$, the Kerr metric describes a rotating black hole.
A Kerr black hole is said to be extremal if the spin parameter admits the maximum permissible value $a=M$. The event
horizon for the extremal black hole is located at $r=M$. If we go past the extremality, i.e., for the spin parameter 
larger than the mass $a>M$, the event horizon is absent and spacetime admits a rotating naked singularity at $r=0$ in the equatorial plane.
The Kerr spacetime admits two Killing vectors, namely a timelike Killing vector $k=\partial_{t}$ and azimuthal Killing vector 
$l=\partial_{\phi}$ that correspond to the time translation and rotational invariance, respectively, which is evident from the 
fact that the Kerr metric \eq{Kerrmetric} is independent of the $t$ and $\phi$.

Consider a massive particle following a geodesic motion on the equatorial plane of the Kerr spacetime. The four-velocity of such a 
particle in the Boyer-Lindquist coordinates is given by 
\begin{eqnarray}
 U^{t} &=& \frac{1}{\Delta}\left(E\left(r^2+a^2+\frac{2Ma^2}{r}\right)-\frac{2Ma}{r}L\right) \ , \nonumber \\ \nonumber \\
 U^{\phi} &=& \frac{1}{\Delta} \left( \frac{2Ma}{r}E+\left(1-\frac{2M}{r}\right)L \right) \ , \nonumber \\ \nonumber \\
 U^{r} &=& \sigma \sqrt{E^2-1+\frac{2M}{r}-\frac{L^2-a^2\left(E^2-1\right)}{r^2}+\frac{2M \left(L-aE\right)^2}{r^3}} \ , \nonumber \\ \nonumber \\
 U^{\theta} &=& 0 \ ,
 \label{velmav}
\end{eqnarray}
where $E=-k \cdot U$ is the conserved energy per unit mass and $L=l \cdot U$ is the conserved angular momentum per unit mass of the particle.
These are the constants of motion associated with the time translation and rotational symmetry. The components of four-velocity \eq{velmav} are obtained by 
solving the equations that define the conserved quantities $E=-k \cdot U$, $L=l \cdot U$ and the normalization condition for velocity $U \cdot U=-1$.
We have $\sigma=\pm 1$ for radially outgoing and ingoing particles, respectively.

We now consider a massless particle following a geodesic motion on the equatorial plane. The four-velocity components of the massless particles moving on the equatorial plane are given by 
\begin{eqnarray}
 U^{t} &=& \frac{1}{\Delta}\left(\left(r^2+a^2+\frac{2Ma^2}{r}\right)-\frac{2Mab}{r}\right) \ , \nonumber \\ \nonumber \\
 U^{\phi} &=& \frac{1}{\Delta} \left( \frac{2Ma}{r}+\left(1-\frac{2M}{r}\right)b \right) \ , \nonumber \\ \nonumber 
\end{eqnarray}
 \begin{eqnarray}
 U^{r} &=& \sigma\sqrt{1-\frac{\left(b^2-a^2\right)}{r^2}+\frac{2M\left(b-a\right)^2}{r^3}} \nonumber \ , \\ \nonumber \\
 U^{\theta} &=& 0 \ ,
 \label{velmal}
\end{eqnarray}
where $b={L}/{E}$ is an impact parameter with $E$ as conserved energy
and $L$ as angular momentum of the massless particle, which are the constants of motion associated with time translational and 
rotational invariance of the Kerr metric, respectively. The components for four-velocity are obtained 
by solving the equations for the conserved quantities $E=-k \cdot U$, $L=l \cdot U$ and the normalization condition for velocity 
of massless particle $U \cdot U=0$ and then changing the parametrization 
$\lambda \rightarrow E\lambda$. Note that $E$ and $L$ are conserved energy and angular momentum per unit mass for massive particle, respectively, 
whereas they also stand for conserved energy and angular momentum for massless particle.

\subsection{LNRF}
One of the most remarkable features of the Kerr metric is the phenomenon of frame-dragging, i.e., spacetime drags 
geodesics sideways in the azimuthal direction in the same direction as that of the rotation. It can be inferred from  Eqs.~\eq{velmav} and \eq{velmal}
as even the particle with the zero angular momentum acquires positive angular velocity when it falls radially inwards.

A locally non-rotating frame (LNRF) is the tetrad associated with an observer who executes a circular
motion at a constant radial coordinate with the frequency associated with frame-dragging. 
Restricting to the equatorial plane, the components of
LNRF basis one-forms are given by
\begin{eqnarray}
e_{\mu}^{(t)} &~:~& e_{t}^{(t)}=\sqrt{\frac{\Delta}{(r^2+a^2+\frac{2M}{r}a^2)}} \ ; e_{\phi}^{(t)}=0 \ ; e_{r}^{(t)}=0 \ ; e_{\theta}^{(t)}=0 \ , \nonumber \\ \nonumber \\
e_{\mu}^{(\phi)} & ~:~& e_{t}^{(\phi)}=-\frac{\frac{2Ma}{r}}{\sqrt{\left(r^2+a^2+\frac{2Ma^2}{r}\right)}} \ ; e_{\phi}^{(\phi)}=\sqrt{\left(r^2+a^2+\frac{2Ma^2}{r}\right)} \ ; e_{r}^{(\phi)}=0 \ ; e_{\theta}^{(\phi)}=0 \ , \nonumber \\ \nonumber \\
e_{\mu}^{(r)} & ~:~& e_{t}^{(r)}=0 \ ; e_{\phi}^{(r)}=0 \ ; e_{r}^{(r)}=\frac{r}{\sqrt{\Delta}} \ ; e_{\theta}^{(r)}=0 \ , \nonumber \\ \nonumber \\
e_{\mu}^{(\theta)} & ~:~& e_{t}^{(\theta)}=0 \ ; e_{\phi}^{(\theta)}=0 \ ; e_{r}^{(\theta)}=0 \ ; e_{\theta}^{(\theta)}=\sqrt{\left(r^2+a^2+\frac{2Ma^2}{r}\right)} \ .
\label{lnrf}
\end{eqnarray}

We now make a transition to the LNRF. We intend to work in the center-of-mass tetrad as it makes computation of the spectrum 
and escape fraction easier. Since the two tetards are related to each other by a Lorentz transformation it would be convenient to 
make a transition to the auxiliary intermediate tetrad from the Boyer-Lindquist coordinates and then figure out the requisite 
Lorentz transformation that would take us from intermediate tetrad to the center of mass frame. We choose auxiliary intermediate 
tetrad to be the locally non-rotating frame \eq{lnrf}. Components of an arbitrarily vector $V$ transform in the following way
as we make a transition from Boyer-Lindquist to LNRF
\begin{equation}
V^{(\mu)}_{LNRF}=e_{\nu}^{(\mu)}V^{\nu} \ .
\label{bltolnrf}
\end{equation}
Thus , Eqs.~\eq{velmav}, \eq{lnrf} and \eq{bltolnrf} , the components of velocity of the massive particle in LNRF are given by 
\begin{eqnarray}
  U^{(t)}_{LNRF} &=& \frac{1}{\sqrt{\Delta}}\frac{\left(E\left(r^2+a^2+\frac{2Ma^2}{r}\right)
  -\frac{2Ma}{r}L\right)}{\sqrt{\left(r^2+a^2+\frac{2Ma^2}{r}\right)}}   \ , \nonumber \\ \nonumber \\
  U^{(\phi)}_{LNRF} &=& \frac{L}{\sqrt{\left(r^2+a^2+\frac{2Ma^2}{r}\right)}} \ , \nonumber \\ \nonumber \\
   U^{(r)}_{LNRF} &=& \sigma \frac{r}{\sqrt{\Delta}} \sqrt{E^2-1+\frac{2M}{r}-\frac{L^2-a^2\left(E^2-1\right)}{r^2}+\frac{2M \left(L-aE\right)^2}{r^3}} \ ,\nonumber \\ \nonumber \\
 U^{(\theta)}_{LNRF} &=&  0 \ ,
 \label{lnrfmav}
\end{eqnarray}
and components for the velocity of the massless particle obtained from Eqs.~\eq{velmal},~\eq{lnrf} and \eq{bltolnrf} are given by 
\begin{eqnarray}
  U^{(t)}_{LNRF} &=& \frac{1}{\sqrt{\Delta}}\frac{\left(\left(r^2+a^2+\frac{2Ma^2}{r}\right)
  -\frac{2Mab}{r}\right)}{\sqrt{\left(r^2+a^2+\frac{2Ma^2}{r}\right)}}  \ , \nonumber \\ \nonumber \\
  U^{(\phi)}_{LNRF} &=& \frac{b}{\sqrt{\left(r^2+a^2+\frac{2Ma^2}{r}\right)}} \ ,\nonumber \\ \nonumber \\
   U^{(r)}_{LNRF} &=& \sigma \frac{r}{\sqrt{\Delta}} \sqrt{1-\frac{\left(b^2-a^2\right)}{r^2}+\frac{2M\left(b-a\right)^2}{r^3}} \ , \nonumber  \\ \nonumber \\
 U^{(\theta)}_{LNRF} &=&  0 \ .
 \label{lnrfmal}
\end{eqnarray}

We consider a collision between two identical massive particles that follow a geodesic motion of the equatorial plane
each with mass $m$ at a radial coordinate $r$. Let the conserved 
energy per unit mass and angular momentum per unit mass for two particles be $E_{i}$ and $L_{i}$ with $i=1,2$, respectively.
It is a priori 
unspecified whether the particles are radially ingoing or outgoing $\sigma_i=\pm$. As we describe later we choose
one of the colliding particles to be radially ingoing and other particle to be radially outgoing. 
The components of the velocity of the colliding particles in the LNRF can be 
obtained from Eq.~\eq{lnrfmav} simply by putting subscript $i$ on various relevant quantities as $U^{(\mu)}_{i,LNRF}$, $E_{i}$, $L_{i}$ and $\sigma_{i}$.

We find it convenient to define the quantities $A$, $B$ and $C$,
which can be interpreted as non-zero components of the net velocity of the two colliding particles 
in LNRF. They would appear in this paper multiple times at various places. While $A$ is always positive for future pointing velocity vectors,
$B$ and $C$ can be positive or negative. Results can be qualitatively different depending on whether $B$ and $C$ are positive or negative.
We analyze these cases separately.

From Eq.~\eq{lnrfmav} we write the expressions for $A$, $B$ and $C$:
\begin{eqnarray}
 A &=& U^{(t)}_{1,LNRF}+U^{(t)}_{2,LNRF} \nonumber \\ 
   &=&  \frac{1}{\sqrt{\Delta}}\frac{\left(E_{1}\left(r^2+a^2+\frac{2Ma^2}{r}\right)
  -\frac{2Ma}{r}L_{1}\right)}{\sqrt{\left(r^2+a^2+\frac{2Ma^2}{r}\right)}}
  + \frac{1}{\sqrt{\Delta}}\frac{\left(E_{2}\left(r^2+a^2+\frac{2Ma^2}{r}\right)
  -\frac{2Ma}{r}L_{2}\right)}{\sqrt{\left(r^2+a^2+\frac{2Ma^2}{r}\right)}} \ , \nonumber \\ \nonumber \\
 B &=&  U^{(r)}_{1,LNRF}+U^{(r)}_{2,LNRF} \nonumber \\ 
   &=& \sigma_{1} \frac{r}{\sqrt{\Delta}} \sqrt{E_{1}^2-1+\frac{2M}{r}-\frac{L_{1}^2-a^2\left(E_{1}^2-1\right)}{r^2}+\frac{2M \left(L_{1}-aE_{1}\right)^2}{r^3}} \nonumber \\
         &&+ \sigma_{2} \frac{r}{\sqrt{\Delta}} \sqrt{E_{2}^2-1+\frac{2M}{r}-\frac{L_{2}^2-a^2\left(E_{2}^2-1\right)}{r^2}+\frac{2M \left(L_{2}-aE_{2}\right)^2}{r^3}} \ , \nonumber \\ \nonumber \\
C &=& U^{(\phi)}_{1,LNRF}+U^{(\phi)}_{2,LNRF}  \nonumber \\ 
  &=& \frac{L_{1}}{\sqrt{\left(r^2+a^2+\frac{2Ma^2}{r}\right)}} +\frac{L_{2}}{\sqrt{\left(r^2+a^2+\frac{2Ma^2}{r}\right)}} \ .
\label{ABC}
\end{eqnarray}
If $B>0$, then the net radial velocity of the two particles in LNRF is positive, which implies the center of mass of the two colliding 
particles moves in the radially outward direction from the point of view of LNRF frame as well as Boyer-Lindquist coordinate system
since the sign of the radial velocity does not change as one makes a transition from the Boyer-Lindquist to LNRF. 
When $B<0$, the center of mass travels in the radially inward direction and when $B=0$, the center of mass does not move along the radial direction. 
The sign of $C$ depicts whether the center of mass is co-rotating or counter-rotating with respect to the spacetime rotation.

\subsection{Center-of-mass frame}

We now make a transition to the center-of-mass frame. It can be achieved by means of an appropriate Lorentz transformation relating 
the LNRF and the center of mass frame. The Lorentz transformation is split into two parts. A rotation that orients the spatial part of the 
net velocity of the two colliding particles in the radial direction and a boost in the radial direction that kills
the radial component of the net velocity. In the center-of-mass frame the net velocity of the two particles has all spatial components zero.

We now write down the transformation that relates the components of any arbitrary vector $V$ in the LNRF to the components
in the center-of-mass frame
\begin{equation}
 V^{(\mu)}_{cm}= \Lambda^{\mu}_{boost \ \nu}\ \Lambda^{\nu}_{rot \ \sigma} \ V_{LNRF}^{(\sigma)} =\Lambda^{\mu}_{boost \ \nu}\ \Lambda^{\nu}_{rot \ \sigma} \ e_{\delta}^{(\sigma)} \ V^{\delta} \ ,
 \label{bltocm}
\end{equation}
where the transformation $\Lambda^{\nu}_{rot \ \sigma}$ appearing in the equation above implements following rotation:
\begin{eqnarray}
 V^{(t)}_{rot} &=&  V^{(t)}_{LNRF} \ ,\nonumber \\ 
 V^{(r)}_{rot} &=& \frac{B}{\sqrt{B^2+C^2}} \ V^{(r)}_{LNRF} + \frac{C}{\sqrt{B^2+C^2}} \ V^{(\phi)}_{LNRF}  \ , \nonumber \\
 V^{(\phi)}_{rot} &=& -\frac{C}{\sqrt{B^2+C^2}} \ V^{(r)}_{LNRF} + \frac{B}{\sqrt{B^2+C^2}} \ V^{(\phi)}_{LNRF}  \ , \nonumber \\
 V^{(\theta)}_{rot} &=&  V^{(\theta)}_{LNRF} \ ,
 \label{rot}
\end{eqnarray}
and the transformation $ \Lambda^{\mu}_{boost \ \nu}$ implements the following boost:
\begin{eqnarray}
 V^{(t)}_{cm} &=& \frac{A}{\sqrt{A^2-B^2-C^2}} \ V^{(t)}_{rot} - \frac{\sqrt{B^2+C^2}}{\sqrt{A^2-B^2-C^2}} \ V^{(r)}_{rot} \ , \nonumber \\ 
 V^{(r)}_{cm} &=& -\frac{\sqrt{B^2+C^2}}{\sqrt{A^2-B^2-C^2}} \ \ V^{(t)}_{rot} + \frac{A}{\sqrt{A^2-B^2-C^2}} \ V^{(r)}_{rot} \ , \nonumber \\
 V^{(\phi)}_{cm} &=&  V^{(\phi)}_{rot}  \ , \nonumber \\
 V^{(\theta)}_{cm} &=&  V^{(\theta)}_{rot} \ .
 \label{boost}
\end{eqnarray}

We now write down the components of velocity of the massive particle in the center of mass frame. 
From Eqs.~\eq{lnrfmav},~\eq{bltocm},~\eq{rot} and \eq{boost} we obtain
\begin{eqnarray}
 U^{(t)}_{cm} &=& \frac{A}{\sqrt{A^2-B^2-C^2}} \ \frac{1}{\sqrt{\Delta}}\frac{\left(\left(r^2+a^2+\frac{2Ma^2}{r}\right)E-\frac{2Ma}{r}L\right)}{\sqrt{\left(r^2+a^2+\frac{2Ma^2}{r}\right)}} 
 - \frac{C}{\sqrt{A^2-B^2-C^2}} \ \frac{L}{\sqrt{\left(r^2+a^2+\frac{2Ma^2}{r}\right)}} 
 \nonumber \\
 &&- \frac{B}{\sqrt{A^2-B^2-C^2}} \sigma \frac{r}{\sqrt{\Delta}}\sqrt{E^2-1+\frac{2M}{r}-\frac{L^2-a^2\left(E^2-1\right)}{r^2}+\frac{2M \left(L-aE\right)^2}{r^3}} \ , 
 \nonumber  
\end{eqnarray}
\begin{eqnarray}
 U^{(r)}_{cm} &=& -\frac{\sqrt{B^2+C^2}}{\sqrt{A^2-B^2-C^2}} \ \frac{1}{\sqrt{\Delta}}\frac{\left(\left(r^2+a^2+\frac{2Ma^2}{r}\right)E-\frac{2Ma}{r}L\right)}{\sqrt{\left(r^2+a^2+\frac{2Ma^2}{r}\right)}} \nonumber \\  
  &&+ \frac{AC}{\sqrt{A^2-B^2-C^2}\sqrt{B^2+C^2}} \ \frac{L}{\sqrt{\left(r^2+a^2+\frac{2Ma^2}{r}\right)}}
 \nonumber \\
 &&+ \frac{AB}{\sqrt{A^2-B^2-C^2}\sqrt{B^2+C^2}} \sigma \frac{r}{\sqrt{\Delta}}\sqrt{E^2-1+\frac{2M}{r}-\frac{L^2-a^2\left(E^2-1\right)}{r^2}+\frac{2M \left(L-aE\right)^2}{r^3}}  \ ,\nonumber \\ 
  \nonumber \\    
 U^{(\phi)}_{cm} &=& -\frac{C}{\sqrt{B^2+C^2}} \ \sigma \frac{r}{\sqrt{\Delta}} \sqrt{E^2-1+\frac{2M}{r}-\frac{L^2-a^2\left(E^2-1\right)}{r^2}+\frac{2M \left(L-aE\right)^2}{r^3}} \nonumber \\
 &&+ \frac{B}{\sqrt{B^2+C^2}} \ \frac{L}{\sqrt{\left(r^2+a^2+\frac{2Ma^2}{r}\right)}} \ , \nonumber \\
 \nonumber \\
 U^{(\theta)}_{cm} &=& 0 \ .
 \label{cmmav}
\end{eqnarray}
The components of the velocity for the massless particle in the center-of-mass frame can be obtained 
from Eqs.~\eq{lnrfmal},~\eq{bltocm},~\eq{rot}
and \eq{boost} and are given by
\begin{eqnarray}
 U^{(t)}_{cm} &=& \frac{A}{\sqrt{A^2-B^2-C^2}} \ \frac{1}{\sqrt{\Delta}}\frac{\left(\left(r^2+a^2+\frac{2Ma^2}{r}\right)-\frac{2Mab}{r}\right)}{\sqrt{\left(r^2+a^2+\frac{2Ma^2}{r}\right)}} 
 - \frac{C}{\sqrt{A^2-B^2-C^2}} \ \frac{b}{\sqrt{\left(r^2+a^2+\frac{2Ma^2}{r}\right)}}
 \nonumber \\
 &&- \frac{B}{\sqrt{A^2-B^2-C^2}} \sigma \frac{r}{\sqrt{\Delta}}\sqrt{1-\frac{\left(b^2-a^2\right)}{r^2}+\frac{2M\left(b-a\right)^2}{r^3}} \ ,\nonumber \\ 
  \nonumber \\    
 U^{(r)}_{cm} &=& -\frac{\sqrt{B^2+C^2}}{\sqrt{A^2-B^2-C^2}} \ \frac{1}{\sqrt{\Delta}}\frac{\left(\left(r^2+a^2+\frac{2Ma^2}{r}\right)-\frac{2Mab}{r}\right)}{\sqrt{\left(r^2+a^2+\frac{2Ma^2}{r}\right)}} \nonumber \\  
  &&+ \frac{AC}{\sqrt{A^2-B^2-C^2}\sqrt{B^2+C^2}} \ \frac{b}{\sqrt{\left(r^2+a^2+\frac{2Ma^2}{r}\right)}} 
 \nonumber \\
 &&+ \frac{AB}{\sqrt{A^2-B^2-C^2}\sqrt{B^2+C^2}} \sigma \frac{r}{\sqrt{\Delta}}\sqrt{1-\frac{\left(b^2-a^2\right)}{r^2}+\frac{2M\left(b-a\right)^2}{r^3}}  \ ,\nonumber \\ 
 \nonumber \\    
U^{(\phi)}_{cm} &=& -\frac{C}{\sqrt{B^2+C^2}} \ \sigma \frac{r}{\sqrt{\Delta}} \sqrt{1-\frac{\left(b^2-a^2\right)}{r^2}+\frac{2M\left(b-a\right)^2}{r^3}} \nonumber \\
 &&+ \frac{B}{\sqrt{B^2+C^2}} \ \frac{b}{\sqrt{\left(r^2+a^2+\frac{2Ma^2}{r}\right)}}  \ , \nonumber  \\ 
 U^{(\theta)}_{cm} &=& 0 \ .
 \label{cmmal}
\end{eqnarray}

\subsection{Collision as seen from the center-of-mass frame}

We now provide a description of the collision event in the center-of-mass frame.
As stated earlier we choose to work in the center-of-mass frame because the
differential cross section of the underlying particle physics process is computed in the center-of-mass frame. It tells us
whether the emission of collision products is isotropic or anisotropic in the center-of-mass frame which is the crucial 
ingredient required to infer the shape of the energy distribution function of the particles escaping to infinity. Thus,
from the point of view of calculation of the observables, it is convenient to work in the center-of-mass frame.

As mentioned earlier, we consider collision between two identical massive particles which we refer to as particle $1$ and particle $2$. 
The colliding particles travel in the opposite directions in the center-of-mass frame with equal and opposite velocity. This can be inferred easily by computing 
the components of net velocity of the colliding particles in the center-of-mass frame. From Eqs.~\eq{ABC} and \eq{cmmav} we obtain
\begin{eqnarray}
 U^{(\alpha)}_{1,cm}+U^{(\alpha)}_{2,cm}=\sqrt{A^2-B^2-C^2}(1,0,0,0).
  \label{umastot}
\end{eqnarray}
The spatial components of net velocity are zero. The time component of the net velocity 
yields the center-of-mass energy of collision  $E_{cm}$
when multiplied by the mass $m$ of the colliding particles
\begin{equation}
 E_{cm}=m \left(U^{(t)}_{1,cm}+ U^{(t)}_{2,cm}\right) = m\sqrt{A^2-B^2-C^2} \ .
\label{ecm}
\end{equation}
We assume that two massless partciles are produced in the collision which we refer to as particle $3$ and particle $4$.
In this paper we focus on the massless particles that move only along the equatorial plane.
The most general expression for the components of the momenta of the two massless particles $P_{3}$ and $P_{4}$ 
moving on the equatorial plane can be written as 
\begin{eqnarray}
 P^{\mu}_{3,cm}= m\frac{\sqrt{A^2-B^2-C^2}}{2}~\left(1,\cos \alpha,\sin \alpha,0\right) \ ,
\label{p3}
\end{eqnarray}
and 
\begin{eqnarray}
 P^{\mu}_{4,cm}= m\frac{\sqrt{A^2-B^2-C^2}}{2}~\left(1,-\cos \alpha,-\sin \alpha,0\right) \ ,
\label{p4}
\end{eqnarray}
respectively. It can be easily verified from the expressions above that $P_3$ and $P_4$ are consistent with energy-momentum 
conservation $ P^{(\mu)}_{3,cm}+ P^{(\mu)}_{3,cm} = m\left(  U^{(\mu)}_{1,cm}+ U^{(\mu)}_{2,cm} \right)$ and are null vectors
$\eta_{\mu\nu}P^{(\mu)}_{3,cm} P^{(\nu)}_{3,cm}=\eta_{\mu\nu} P^{(\mu)}_{4,cm} P^{(\nu)}_{4,cm}=0$. Further it is quite clear from 
the transformation Eqs.~\eq{bltocm},~\eq{lnrf},~\eq{rot} and \eq{boost} from the 
Boyer-Lindquist coordinate system to LNRF and Eqs.~\eq{p3} and \eq{p4} that $P^{(\theta)}_{3}= P^{(\theta)}_{4} = 0$.
Thus, particles move on the equatorial plane. In the center-of-mass frame particles move in the $\hat{r}-\hat{\phi}$ plane. The parameter 
$\alpha$ which appears in Eqs.~\eq{p3} and \eq{p4} can be interpreted as the angle between the direction in which particle $3$ travels and $\hat{r}$ direction.
Whereas particle $4$ travels along the direction which makes angle $\pi+\alpha$ with $\hat{r}$. The particles travel along opposite directions 
with equal and opposite momenta in the center of mass frame.

We now compute the conserved energies and angular momenta of the two massless particles 
produced in the collision. We write down the components of the
timelike Killing vector $k$ and azimuthal Killing vector $l$ in the center-of-mass frame. Their components
in the Boyer-Lindquist coordinate system are 
\begin{eqnarray}
 k^{\mu}&=&(1,0,0,0) \ , \nonumber \\
 l^{\mu}&=&(0,0,1,0) \ .
 \label{klbl}
\end{eqnarray}

From Eqs.~\eq{klbl},~\eq{bltocm},~\eq{lnrf},~\eq{rot} and \eq{boost}, the components of the timelike Killing vector in the 
center-of-mass frame can be written as
\begin{eqnarray}
 k^{(t)}_{cm}&=& \frac{A}{\sqrt{A^2-B^2-C^2}}\sqrt{\frac{\Delta}{\left(r^2+a^2+\frac{2Ma^2}{r}\right)}}+\frac{C}{\sqrt{A^2-B^2-C^2}}\frac{\frac{2Ma}{r}}{\sqrt{r^2+a^2+\frac{2Ma^2}{r}}} \ , \nonumber \\
 k^{(r)}_{cm}&=& -\frac{\sqrt{B^2+C^2}}{\sqrt{A^2-B^2-C^2}}\sqrt{\frac{\Delta}{\left(r^2+a^2+\frac{2Ma^2}{r}\right)}} - \frac{AC}{\sqrt{A^2-B^2-C^2}}\frac{\frac{2Ma}{r}}{\sqrt{B^2+C^2}} \ , \nonumber \\
 k^{(\phi)}_{cm}&=& -\frac{B}{\sqrt{B^2+C^2}} \frac{\frac{2Ma}{r}}{\sqrt{\left(r^2+a^2+\frac{2Ma^2}{r}\right)}} \ , \nonumber \\
 k^{(\theta)}_{cm}&=&0 \ ,
 \label{kcm}
\end{eqnarray}
and the components of azimuthal Killing vector are given by 
\begin{eqnarray}
 l^{(t)}_{cm}&=& -\frac{C}{\sqrt{A^2-B^2-C^2}}\sqrt{\frac{\Delta}{\left(r^2+a^2+\frac{2Ma^2}{r}\right)}}  \ ,\nonumber \\
 l^{(r)}_{cm}&=& \frac{AC}{\sqrt{A^2-B^2-C^2}}\frac{\sqrt{\left(r^2+a^2+\frac{2Ma^2}{r}\right)}}{\sqrt{B^2+C^2}} \ , \nonumber \\
 l^{(\phi)}_{cm}&=& \frac{B}{\sqrt{B^2+C^2}} \sqrt{\left(r^2+a^2+\frac{2Ma^2}{r}\right)} \ , \nonumber \\
 l^{(\theta)}_{cm}&=&0 \ .
 \label{lcm}
\end{eqnarray}
The conserved energies $E_i$ and angular momenta $L_i$ are obtained by taking the inner product of the timelike and azimuthal Killing vectors 
with the momentum vectors of the two particles produced in the collision. The components of all the relevant quantities are written 
down in the center-of-mass frame.  
From Eqs.~\eq{p3},~\eq{p4},~\eq{kcm} and \eq{lcm}, the conserved energies of the massless particles $E_3$ and $E_4$ are given by
\begin{eqnarray}
 E_{3} &=& -\eta_{\mu\nu}P^{(\mu)}_{3,cm}k^{(\nu)}_{cm} \nonumber \\
       &=& \frac{m}{2}\frac{\sqrt{\Delta}A+\frac{2Ma}{r}C}{\sqrt{\left(r^2+a^2+\frac{2Ma^2}{r}\right)}}
       + \frac{m}{2}\frac{\sqrt{\Delta}\sqrt{B^2+C^2}+\frac{2Ma}{r}\frac{AC}{\sqrt{B^2+C^2}}}{\sqrt{\left(r^2+a^2+\frac{2Ma^2}{r}\right)}}\cos\alpha \nonumber \\
       &&+ \frac{m}{2}
	\frac{\frac{2Ma}{r}}{\sqrt{\left(r^2+a^2+\frac{2Ma^2}{r}\right)}}\frac{B}{\sqrt{B^2+C^2}}\sqrt{A^2-B^2-C^2}
	\sin \alpha \ ,
\label{E3}
\end{eqnarray}
and
\begin{eqnarray}
 E_{4} &=& -\eta_{\mu\nu}P^{(\mu)}_{4,cm}k^{(\nu)}_{cm} \nonumber \\
       &=& \frac{m}{2}\frac{\sqrt{\Delta}A+\frac{2Ma}{r}C}{\sqrt{\left(r^2+a^2+\frac{2Ma^2}{r}\right)}}
       - \frac{m}{2}\frac{\sqrt{\Delta}\sqrt{B^2+C^2}+\frac{2Ma}{r}\frac{AC}{\sqrt{B^2+C^2}}}{\sqrt{\left(r^2+a^2+\frac{2Ma^2}{r}\right)}}\cos\alpha \nonumber \\
       &&- \frac{m}{2} \frac{\frac{2Ma}{r}}{\sqrt{\left(r^2+a^2+\frac{2Ma^2}{r}\right)}}\frac{B}{\sqrt{B^2+C^2}}\sqrt{A^2-B^2-C^2} \sin \alpha \ ,
\label{E4}
\end{eqnarray} 
respectively, and angular momenta $L_3$ and $L_4$ can be written as 
\begin{eqnarray}
 L_{3} &=& \eta_{\mu\nu}P^{(\mu)}_{3,cm}l^{(\nu)}_{cm} \nonumber \\
       &=& \frac{m}{2}\sqrt{\left(r^2+a^2+\frac{2Ma^2}{r}\right)} C + \frac{m}{2}\sqrt{\left(r^2+a^2+\frac{2Ma^2}{r}\right)} \frac{AC}{\sqrt{B^2+C^2}} \cos\alpha \nonumber \\
       &&+   \frac{m}{2}\sqrt{\left(r^2+a^2+\frac{2Ma^2}{r}\right)}
	\frac{B}{\sqrt{B^2+C^2}}\sqrt{A^2-B^2-C^2} \sin\alpha \ ,
\label{L3}
\end{eqnarray}
and
\begin{eqnarray}
 L_{4} &=& \eta_{\mu\nu}P^{(\mu)}_{3,cm}l^{(\nu)}_{cm} \nonumber \\
       &=& \frac{m}{2}\sqrt{\left(r^2+a^2+\frac{2Ma^2}{r}\right)} C - \frac{m}{2}\sqrt{\left(r^2+a^2+\frac{2Ma^2}{r}\right)} \frac{AC}{\sqrt{B^2+C^2}} \cos\alpha \nonumber \\
       &&-   \frac{m}{2}\sqrt{\left(r^2+a^2+\frac{2Ma^2}{r}\right)} \frac{B}{\sqrt{B^2+C^2}}\sqrt{A^2-B^2-C^2} \sin\alpha \ ,
\label{L4}
\end{eqnarray}
respectively.

If the massless particle produced in the collision escapes to
infinity, then its conserved energy is the energy of the particle as 
measured by the asymptotic observer. Thus, no further effort is required to infer the energy of the massless particle measured at infinity
than to specify the direction along which it is emitted in the center-of-mass frame. We must determine the directions along which 
the particle must be emitted in the center-of-mass frame so that it escapes to infinity. 
Without the loss of generality we label the particle which escapes to infinity as particle $3$. 

\begin{figure}
\begin{center}
\includegraphics[width=0.9\textwidth]{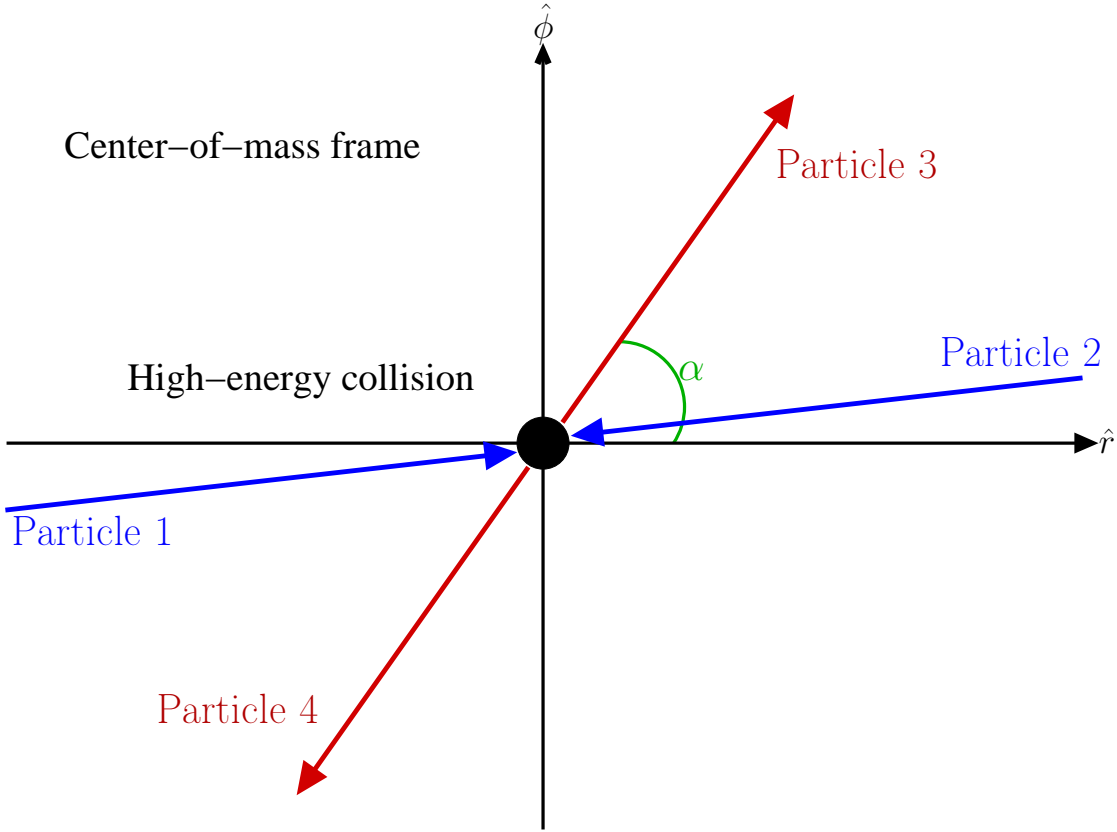}
\caption{ Collision as seen from the center-of-mass tetrad. Two massive particles, namely 
particle $1$ and particle $2$, collide and produce two massless particles, namely particle $3$ and particle $4$. 
All particles travel in $\hat{r}-\hat{\phi}$ plane in the center-of-mass frame. The
colliding particles as well as collision products travel in the 
opposite direction with equal and opposite momenta. Particle $3$ travels along the direction 
which makes angle $\alpha$ with $\hat{r}$.
}
\label{collision}
\end{center}
\end{figure}

\subsection{Conditions for the massless particle to escape to infinity}

\begin{figure}
\begin{center}
\includegraphics[width=0.9\textwidth]{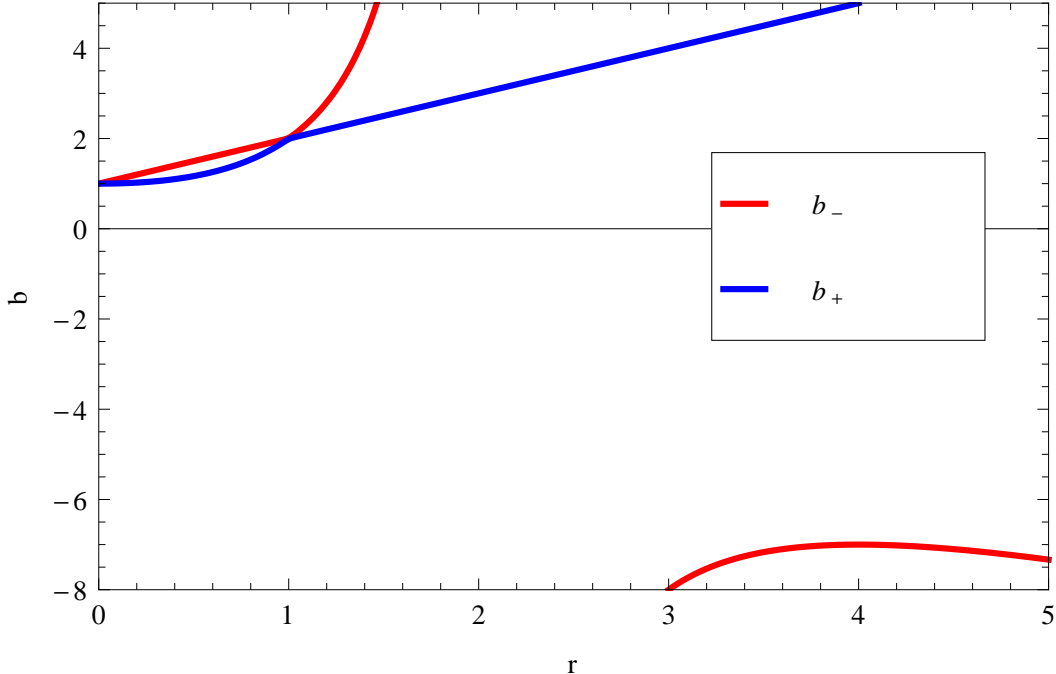}
\caption{The impact parameter $b(r)$ required for the massless particles to admit a turning point at a radial coordinate $r$
is plotted here. Both $b$ and $r$ are expressed in units of $M$. $b$ admits two branches, $b_{-}$ denoated by red curve and 
$b_{+}$ denoted by blue curve. The Kerr spin parameter is taken to be $a=1.0001M$.
}
\label{figbr}
\end{center}
\end{figure}

We now derive the conditions necessary for the massless particle produced in the collision to escape to infinity. 
Consider a massless particle with an impact parameter $b$ following a geodesic motion on the equatorial plane of the Kerr spacetime. The radial component of velocity 
in the Boyer-Lindquist coordinate system can be read off from Eq.~\eq{velmal} as
\begin{equation}
  U^{r} =\pm \sqrt{1-\frac{\left(b^2-a^2\right)}{r^2}+\frac{2M\left(b-a\right)^2}{r^3}} \ .
\end{equation}
The value that the impact parameter $b(r)$ must take if it is to admit a turning point at a radial coordinate $r$, i.e. for $U^{r}=0$ 
is given by the expression
\begin{equation}
 b_{\pm}(r)=\frac{-2Ma \pm r\sqrt{\Delta}}{\left(r-2M\right)} \ .
 \label{br}
\end{equation} 
The expression above contains the factor $\sqrt{\Delta}$. Analysis for the Kerr black hole and the overspinning Kerr geometry 
must be carried out
separately since for the black hole $\Delta=0$ at the horizon $r=r_h$ and the region below horizon $r<r_{h}$ is irrelevant for the discussion,
whereas in case of the naked singularity $\Delta$ remains positive throughout the spacetime. Here we focus on the ovespinning 
Kerr geometry since it is relevant from the point of view of results discussed in this paper. An analysis for the Kerr black hole was carried 
out in \cite{Harada1},\cite{BSW2}.

We first analyze the behavior of $b_{+}(r)$. It is a monotonically increasing function which is always positive.
At the location of the singularity, i.e., at $r=0$,
it takes a value $b_{+}(r=0)=a$. At $r=2M$ both the denominator and numerator vanish in the expression for $b_{+}$. However one
can take an appropriate limit and obtain $ b_{+}(r=2M)=a+({2M^2}/{a})$. As we approach infinity it shows a
divergence $ b_{+}(r \rightarrow \infty) \rightarrow \infty$.

We now analyze the behavior of $b_{-}(r)$. When $r<2M$, $b_{-}(r)$ is always larger than $b_{+}(r)$, i.e., $b_{-}(r)>b_{+}(r)$.
At the location of the singularity $r=0$, it takes value $b_{-}(r=0)=a$. When we approach $r=2M$ from left, it 
goes to plus infinity $b_{-}(r\rightarrow 2M^{-}) \rightarrow +\infty$. When we approach $r=2M$ from right, 
$b_{-}$ goes to negative infinity. For $r>2M$, $b_{-}(r)$ is always negative.
The radial coordinate $r_{m}$ at which $b_{-}$ admits maximum and its maximum value are given by 
\begin{equation}
r_{m}=\frac{M^{\frac{1}{3}}\left(M^{\frac{2}{3}}+\left(a+\sqrt{a^2-M^2}\right)^{\frac{2}{3}}\right)^2}{\left(a+\sqrt{a^2-M^2}\right)^{\frac{2}{3}}}
\ ; \ 
b_{-}(r=r_{m})=a-\frac{\left(M^{\frac{2}{3}}+\left(a+\sqrt{a^2-M^2}\right)^{\frac{2}{3}}\right)^3}{\left(a+\sqrt{a^2-M^2}\right)} \ .
\label{rmbm}
\end{equation}
As we approach infinity it goes to negative infinity $ b_{-}(r\rightarrow \infty) \rightarrow -\infty$. Behavior of 
$b_{+}(r)$ and $b_{-}(r)$ is as shown in Fig.~\ref{figbr}.

Conditions that must be imposed on the impact parameter of the massless particle created in the collision so that it escapes to infinity 
are different depending on whether the collision takes place at the location with $r < r_{m}$ or $r > r_{m}$. 

If the collision occurs at $r<r_{m}$ and if the massless particle travels radially inwards, i.e., $\sigma=-1$,
then its impact parameter must be in the following range for it to admit a turning point at the lower radius and 
escape to infinity:
\begin{equation}
 r<r_{m} \ , \ \sigma=-1 \  \implies \ b \in \left(a \ , b_{+}(r)\right) \ .
 \label{blin}
\end{equation}
If the massless particle travels radially outwards, i.e.,
$\sigma=+1$, then its impact parameter must be in the following
range for it to escape:
\begin{equation}
 r<r_{m}\ , \ \sigma=+1 \ \implies \ b \in \left( b_{-}\left(r_{m}\right) , b_{+}(r)\right) \ .
 \label{blout}
\end{equation}

If the collision occurs at $r>r_{m}$ and if the massless particle travels radially inwards, i.e., $\sigma=-1$,
then its impact parameter must be in the following range for it to turn back at the lower radius and 
escape to infinity:
\begin{equation}
 r>r_{m} \ , \ \sigma=-1 \  \implies \ b \in \left(a \ , \ b_{+}(r)\right) \cup 
 \left( b_{-}(r) \ , \ b_{-}\left(r_m\right) \right) \ .
 \label{bgin}
\end{equation}
If the massless particle travels radially outwards $\sigma=+1$, then its impact parameter must be in the following range for it to escape:
\begin{equation}
 r>r_{m}\ , \ \sigma=+1 \ \implies \ b \in \left( b_{-}(r) \ , \ b_{+}(r)\right) \ .
 \label{bgout}
\end{equation}

If these conditions are met, then the massless particle created in the particle collision will escape to infinity. 

\subsection{Escape cones in the center-of-mass frame}
We derived the conditions that must be imposed on the impact parameter of the massless particle generated 
in the collision for it to escape to infinity. We now translate these conditions into the set of directions 
along which the particle must be emitted in the center-of-mass frame. We compute the escape cones, i.e., the 
permissible angular range for the escape of the massless particle.

The massless particle moves in the $\hat{r}-\hat{\phi}$ plane in the center-of-mass frame.
The nonzero spatial components of the velocity in the center-of-mass frame 
can be read off from Eq.~\eq{cmmal} and are given by 
\begin{eqnarray}
 U^{(r)}_{cm} &=& \frac{-\sqrt{B^2+C^2}}{\sqrt{A^2-B^2-C^2}} \ \frac{1}{\sqrt{\Delta}}\frac{\left(\left(r^2+a^2+\frac{2Ma^2}{r}\right)-\frac{2Mab}{r}\right)}{\sqrt{\left(r^2+a^2+\frac{2Ma^2}{r}\right)}} \nonumber \\
 &&+ \frac{AB}{\sqrt{A^2-B^2-C^2}\sqrt{B^2+C^2}} \sigma \frac{r}{\sqrt{\Delta}}\sqrt{1-\frac{\left(b^2-a^2\right)}{r^2}+\frac{2M\left(b-a\right)^2}{r^3}} \nonumber \\ 
 &&+ \frac{AC}{\sqrt{A^2-B^2-C^2}\sqrt{B^2+C^2}} \ \frac{b}{\sqrt{\left(r^2+a^2+\frac{2Ma^2}{r}\right)}} \ ,  
 \label{umlcmrphi0}
 \end{eqnarray}
 and
 \begin{eqnarray}
 U^{(\phi)}_{cm} &=& -\frac{C}{\sqrt{B^2+C^2}} \ \sigma \frac{r}{\sqrt{\Delta}} \sqrt{1-\frac{\left(b^2-a^2\right)}{r^2}+\frac{2M\left(b-a\right)^2}{r^3}} \nonumber \\
 &&+ \frac{B}{\sqrt{B^2+C^2}} \ \frac{b}{\sqrt{\left(r^2+a^2+\frac{2Ma^2}{r}\right)}} \ . 
 \label{umlcmrphi}
\end{eqnarray}

The angle $\alpha$ subtended by the direction along which the massless particle travels in the center-of-mass frame with respect to $\hat{r}$
can be obtained from the velocity components \eq{umlcmrphi0},~\eq{umlcmrphi} by solving the equations below.
\begin{equation}
\sin \alpha = \frac{ U^{(\phi)}_{cm}}{ \sqrt{U^{(r) \ 2}_{cm}+ U^{(\phi) \ 2}_{cm}}} \ ; \  \cos \alpha = \frac{ U^{(r)}_{cm}}{ \sqrt{U^{(r) \ 2}_{cm}+ U^{(\phi) \ 2}_{cm}}}
 \label{alphamal}
\end{equation}

We define the following critical angles for the radially ingoing particles with critical
impact parameters encountered earlier in Eqs.~\eq{blin} and \eq{bgin} :
\begin{eqnarray}
\alpha_1 &&=  \alpha\left(\sigma=-1 \ , \ r \ , \ b=a\right) \ ,\nonumber \\
\alpha_2 &&= \alpha\left(\sigma=-1 \ , \ r \ , \ b=b_{+}(r)\right) \ , \nonumber \\
\alpha_3 &&= \alpha\left(\sigma=-1 \ , \ r \ , \ b=b_{-}(r)\right) \ ,\nonumber \\
\alpha_4 &&= \alpha\left(\sigma=-1 \ , \ r \ , \ b=b_{-}\left(r_{m}\right)\right) \ . \nonumber \\
 \label{alphaesc+}
\end{eqnarray}
Similarly, we define critical angles for radially outgoing particles with relevant critical impact parameters encountered in Eqs.~\eq{blout} and \eq{bgout}
as follows:
\begin{eqnarray}
\alpha_5 &&= \alpha\left(\sigma=+1 \ , \ r \ , \ b=b_{+}(r)\right) \ , \nonumber \\
\alpha_6 &&= \alpha\left(\sigma=+1 \ , \ r \ , \ b=b_{-}(r)\right) \ , \nonumber \\
\alpha_7 &&= \alpha\left(\sigma=+1 \ , \ r \ , \ b=b_{-}\left(r_{m}\right)\right) \ .
\label{alphaesc-}
\end{eqnarray}

We define 
\begin{equation}
 \left[ \left[ \alpha_i,\alpha_j \right] \right]= \left( min\left(\alpha_i,\alpha_j\right), max\left(\alpha_i,\alpha_j\right) \right)
 \label{defbrac}
\end{equation}

If collision occurs at $r<r_{m}$, then the escape cone $EC_{r<r_{m}}$ for the massless particles can be
obtained from Eqs.~\eq{blin},~\eq{blout},~\eq{alphaesc+} and \eq{alphaesc-}. It is given by 
\begin{equation}
 EC_{r<r_{m}}= \left[\left[\alpha_1,\alpha_2\right]\right]\cup \left[\left[\alpha_5,\alpha_7\right]\right] \ ,
 \label{ecless}
\end{equation}
whereas in the case collision occurs at $r>r_{m}$ then the escape cone $EC_{r>r_{m}}$
obtained from Eqs.~\eq{bgin},~\eq{bgout},~\eq{alphaesc+} and \eq{alphaesc-} is given by 
\begin{equation}
 EC_{r>r_{m}}= \left[\left[\alpha_1,\alpha_2\right]\right]\cup \left[\left[\alpha_3,\alpha_4\right]\right]\cup \left[\left[\alpha_5,\alpha_6\right]\right] \ .
\end{equation}
Thus, we have identified the escape cones within which the massless particle must be emitted in the center-of-mass frame 
so that it escapes to infinity.

\subsection{Collisional Penrose process}

We now determine the conserved energy of the massless particle that is emitted within the escape cone. 
It depends on the direction along which the massless particle is emitted in the center-of-mass frame, i.e., 
it depends on the angle $\alpha$ with $\alpha \in  EC_{r<r_{m}}$ if $r<r_{m}$,
or $\alpha$ with $\alpha \in  EC_{r>r_{m}}$ if $r>r_{m}$. Since the particle within the escape cone reaches infinity, 
the conserved energy of the particle is also its energy $E$ 
measured by asymptotic observer. 

From Eq.~\eq{E3} it is given by the expression 
\begin{eqnarray}
E(\alpha) 
       &=& \frac{m}{2}\frac{\sqrt{\Delta}A+\frac{2Ma}{r}C}{\sqrt{\left(r^2+a^2+\frac{2Ma^2}{r}\right)}}
       + \frac{m}{2}\frac{\sqrt{\Delta}\sqrt{B^2+C^2}+\frac{2Ma}{r}\frac{AC}{\sqrt{B^2+C^2}}}{\sqrt{\left(r^2+a^2+\frac{2Ma^2}{r}\right)}}\cos\alpha \nonumber \\
       &&+ \frac{m}{2} \frac{\frac{2Ma}{r}}{\sqrt{\left(r^2+a^2+\frac{2Ma^2}{r}\right)}}\frac{B}{\sqrt{B^2+C^2}}\sqrt{A^2-B^2-C^2} \sin \alpha \ ,
\label{E}
\end{eqnarray}
whereas the conserved energy of the other particle produced in the collision which is emitted in the opposite direction, 
denoted by $E'$ can be obtained from Eq.~\eq{E4} and is given by
\begin{eqnarray}
E'(\alpha) 
       &=& \frac{m}{2}\frac{\sqrt{\Delta}A+\frac{2Ma}{r}C}{\sqrt{\left(r^2+a^2+\frac{2Ma^2}{r}\right)}}
       - \frac{m}{2}\frac{\sqrt{\Delta}\sqrt{B^2+C^2}+\frac{2Ma}{r}\frac{AC}{\sqrt{B^2+C^2}}}{\sqrt{\left(r^2+a^2+\frac{2Ma^2}{r}\right)}}\cos\alpha \nonumber \\
       &&- \frac{m}{2} \frac{\frac{2Ma}{r}}{\sqrt{\left(r^2+a^2+\frac{2Ma^2}{r}\right)}}\frac{B}{\sqrt{B^2+C^2}}\sqrt{A^2-B^2-C^2} \sin \alpha \ .
\label{E'}
\end{eqnarray}

Because of the phenomenon of frame-dragging, the timelike Killing vector $k=\partial_{t}$ can become a 
spacelike vector and the conserved energy can take a
negative value. For some set of directions $\alpha$, if $E'(\alpha)$ turns out to be negative and then the
energy of the particle escaping to infinity $E(\alpha)$ would be larger than the combined energy of the 
colliding particles and collisional Penrose process would be at work. That is
\begin{equation}
E'(\alpha)<0 \ \implies \ E(\alpha)>m\left(E_1+E_2\right) \ .
\label{cp}
\end{equation}
The efficiency of the collisional Penrose process is defined as the ratio of the energy of the escaping particle 
created in the collision to the combined energy of the colliding particles. 
\begin{equation}
 \eta=\frac{E}{m\left(E_1+E_2\right)}
 \label{effdef}
\end{equation}
It would be possible to generate ultra-high-energy particles, starting from the particles with moderate energies if the efficiency 
of the collisional Penorse process takes a value that is very large, i.e.,
\begin{equation}
 \eta \gg 1 \ .
 \label{uhep}
\end{equation}
This condition is not met in the case of the Kerr black hole as it was demonstrated in \cite{Bejger},\cite{Harada1}. In this paper 
we show that efficiency of collisional Penrose process can be divergent for near-extremal overspinning Kerr geometry.

\subsection{Condition for the divergence of the efficiency of collisional Penrose process}
We now identify the necessary conditions for the divergence of efficiency of the collisional Penrose process
in the Kerr spacetime. First it is necessary for the conserved energy of the massless particle produced 
in the collision to diverge. Secondly this particle must be emitted within the escape cone so that it escapes 
to infinity. 

The conserved energy of particle $3$ is given by the expression as it can be read off from Eq.~\eq{E3}
\begin{eqnarray}
E_3 
       &=& \frac{m}{2}\frac{\sqrt{\Delta}A+\frac{2Ma}{r}C}{\sqrt{\left(r^2+a^2+\frac{2Ma^2}{r}\right)}}
       + \frac{m}{2}\frac{\sqrt{\Delta}\sqrt{B^2+C^2}+\frac{2Ma}{r}\frac{AC}{\sqrt{B^2+C^2}}}{\sqrt{\left(r^2+a^2+\frac{2Ma^2}{r}\right)}}\cos\alpha \nonumber \\
       &&+ \frac{m}{2} \frac{\frac{2Ma}{r}}{\sqrt{\left(r^2+a^2+\frac{2Ma^2}{r}\right)}}\frac{B}{\sqrt{B^2+C^2}}\sqrt{A^2-B^2-C^2} \sin \alpha \ ,
\label{E1}
\end{eqnarray}
where $A$,$B$ and $C$ are given by Eq.~\eq{ABC} as follows:
\begin{eqnarray}
 A &=&  \frac{1}{\sqrt{\Delta}}\frac{\left(E_{1}\left(r^2+a^2+\frac{2Ma^2}{r}\right)
  -\frac{2Ma}{r}L_{1}\right)}{\sqrt{\left(r^2+a^2+\frac{2Ma^2}{r}\right)}}
  + \frac{1}{\sqrt{\Delta}}\frac{\left(E_{2}\left(r^2+a^2+\frac{2Ma^2}{r}\right)
  -\frac{2Ma}{r}L_{2}\right)}{\sqrt{\left(r^2+a^2+\frac{2Ma^2}{r}\right)}} \ , \nonumber \\ \nonumber \\
 B &=& \sigma_{1} \frac{r}{\sqrt{\Delta}} \sqrt{E_{1}^2-1+\frac{2M}{r}-\frac{L_{1}^2-a^2\left(E_{1}^2-1\right)}{r^2}+\frac{2M \left(L_{1}-aE_{1}\right)^2}{r^3}} \nonumber \\
         &&+ \sigma_{2} \frac{r}{\sqrt{\Delta}} \sqrt{E_{2}^2-1+\frac{2M}{r}-\frac{L_{2}^2-a^2\left(E_{2}^2-1\right)}{r^2}+\frac{2M \left(L_{2}-aE_{2}\right)^2}{r^3}} \ ,\nonumber 
\end{eqnarray}   
and
\begin{eqnarray}
C &=& \frac{L_{1}}{\sqrt{\left(r^2+a^2+\frac{2Ma^2}{r}\right)}} +\frac{L_{2}}{\sqrt{\left(r^2+a^2+\frac{2Ma^2}{r}\right)}} \ .
\label{ABC1}
\end{eqnarray}

For the finite values of the radial coordinate $r$ and geodesic parameters $E_1$,$E_2$,$L_1$ and $L_2$, 
it is clear from Eq.~\eq{ABC1}
that $C$ is always finite. Whereas, $A$ and $B$ can potentially diverge when $\Delta \rightarrow 0$, i.e.,
\begin{eqnarray}
 C &=& O(1) \ , \nonumber \\ \Delta \rightarrow 0    &\implies& \ A,B \sim \frac{1}{\sqrt{\Delta}} \rightarrow \infty \ .
 \label{abcdiv}
\end{eqnarray}

Since $A$,$B$ and $C$ are the components of the net velocity of the two colliding particles which is timelike vector, we 
have $A>0$,$~A>|B|$ and $A>|C|$. Further, one can infer from Eq.~\eq{ABC1} that $B$ and $C$ can never be zero together.
Thus, when $A$ is finite,
$E_3$ would be finite. For $E_3$ to diverge, $A$ must necessarily diverge. From Eq.~\eq{abcdiv}, $A$ diverges when $\Delta \rightarrow 0$
and $B$ can also potentially diverge. If both $A$ and $B$ diverge and are almost equal in magnitude, i.e.,
if ${(A-|B|)}/{A}\rightarrow 0^{+}$,
then $E_3$ is finite. For $E_3$ to diverge $A$ must be significantly
larger than $B$, i.e., $ {(A-|B|)}/{A} = O(1)$
. Thus, the requisite conditions for the divergence of $E_3$ are as follows:
\begin{equation}
 A \rightarrow \infty ~~~ \text{and}  ~~~ \frac{A-|B|}{A} = O(1).
 \label{con1}
\end{equation}

Interestingly, if the conditions stated above are met, then the center-of-mass energy of the two colliding particles
also shows divergence. From Eq.~\eq{ecm},~\eq{abcdiv} and \eq{con1} these conditions above are equivalent to 
\begin{equation}
 E_{cm}=m\sqrt{A^2-B^2-C^2} \rightarrow \infty \ .
 \label{ecminfty}
\end{equation}

We obtain a crucial result here, namely, that we must have collision with the divergent center of mass energy if we want 
the efficiency of collisional Penrose process to diverge. Note that it is necessary but not a sufficient condition. 
The conserved energy of the massless particle created in the collision can diverge if the 
center-of-mass energy shows divergence.
Further, the particle with large energy should also escape to infinity 
if the efficiency of collisional process is to show divergence. 

It is possible to arrange for the collisions with the divergent center-of-mass energy 
around the near-extremal Kerr black holes and in the overspinning Kerr geometry \cite{BSW},\cite{Patil1},\cite{Patil2}.
However the efficiency of collisional Penrose process is finite in the case of black holes \cite{Bejger},\cite{Harada1}, since the 
particles with large conserved energies produced in the ultra-high-energy collisions do not escape to infinity, but
eventually enter the black hole. In this paper we show that in the overspinning Kerr geometry, 
particles with divergent conserved energies can also escape to infinity. 
Thus, the efficiency of collisional Penrose process shows divergence.

\subsection{Direction along which the colliding particles travel in center-of-mass frame}
We determine the direction along which the colliding particles travel in the center-of-mass frame. This is needed 
for the calculation of the spectrum of the massless particles escaping to infinity, since the distribution of the emission 
of massless particles in the center-of-mass frame depends on the angle between the direction in which the colliding particles travel and direction 
along which the massless particles are emitted.

The colliding particles travel in $\hat{r}-\hat{\phi}$ plane in the center-of-mass frame. The non-zero spatial 
components of velocity can be obtained from Eq.~\eq{cmmav} by putting subscript $i=1,2$ on relevant quantities and are given by 
\begin{eqnarray}
 U^{(r)}_{i,cm} &=& -\frac{\sqrt{B^2+C^2}}{\sqrt{A^2-B^2-C^2}} \ \frac{1}{\sqrt{\Delta}}\frac{\left(\left(r^2+a^2+\frac{2Ma^2}{r}\right)E_i-\frac{2Ma}{r}L_i\right)}{\sqrt{\left(r^2+a^2+\frac{2Ma^2}{r}\right)}} \nonumber \\  
  &&+ \frac{AC}{\sqrt{A^2-B^2-C^2}\sqrt{B^2+C^2}} \ \frac{L_i}{\sqrt{\left(r^2+a^2+\frac{2Ma^2}{r}\right)}}
 \nonumber \\
 &&+ \frac{AB}{\sqrt{A^2-B^2-C^2}\sqrt{B^2+C^2}} \sigma \frac{r}{\sqrt{\Delta}}\sqrt{E_i^2-1+\frac{2M}{r}-\frac{L_i^2-a^2\left(E_i^2-1\right)}{r^2}+\frac{2M \left(L_i-aE_i\right)^2}{r^3}}  \ , \nonumber 
 \end{eqnarray}
and
\begin{eqnarray}
 U^{(\phi)}_{i,cm} &=& -\frac{C}{\sqrt{B^2+C^2}} \ \sigma \frac{r}{\sqrt{\Delta}} \sqrt{E_i^2-1+\frac{2M}{r}-\frac{L_i^2-a^2\left(E_i^2-1\right)}{r^2}+\frac{2M \left(L_i-aE_i\right)^2}{r^3}} \nonumber \\
 &&+ \frac{B}{\sqrt{B^2+C^2}} \ \frac{L_i}{\sqrt{\left(r^2+a^2+\frac{2Ma^2}{r}\right)}} \ .\nonumber \\
 \label{umas}
\end{eqnarray}
The angle $\alpha_{c,i}$ made by the direction along which the colliding particle travels in the center-of-mass frame can be obtained from
the velocity components Eq.~\eq{umas} by solving equations below
\begin{equation}
\sin \alpha_{c,i} = \frac{ U^{(\phi)}_{i,cm}}{ \sqrt{U^{(r) \ 2}_{i,cm}+ U^{(\phi) \ 2}_{i,cm}}} \ ; \  \cos \alpha_{c,i} = \frac{ U^{(r)}_{i,cm}}{ \sqrt{U^{(r) \ 2}_{i,cm}+ U^{(\phi) \ 2}_{i,cm}}} \ .
 \label{alphamas}
\end{equation}
Since, from Eq.~\eq{umastot} in the center-of-mass frame, we have 
\begin{equation}
  U^{(r)}_{1,cm}+ U^{(r)}_{2,cm}= U^{(\phi)}_{i,cm}+ U^{(\phi)}_{i,cm}=0 \ .
  \label{utot1}
\end{equation}
From Eqs.~\eq{alphamas} and \eq{utot1}, we obtain
\begin{equation}
 \alpha_{c,2}=\pi+\alpha_{c,1}
 \label{eq90}
\end{equation}
i.e. two particles move in the opposite directions.

\subsection{Energy distribution of the massless particles escaping to infinity}

We now compute the energy spectrum of the massless particles produced in the collisions escaping to infinity.
We restrict ourselves to the collisions taking place at the specific radial coordinate since the 
ultra-high-energy collisions take 
place in a very narrow band over radial coordinate \cite{BSW},\cite{Patil1},\cite{Patil2}.
The energy of the massless particle produced in the collision depends on the conserved energies and angular momenta
of the colliding particles $E_1$,$E_2$,$L_1$ and $L_2$, information whether 
particle is radially ingoing or outgoing, i.e., the values of $\sigma_1$ and~$\sigma_2$, and 
the angle $\alpha$ at which the particle is emitted in the center-of-mass frame \eq{E}. A given value of energy $E$ of the massless particle
escaping to infinity can be achieved via different combinations of $E_1,E_2,L_1,L_2,\sigma_1,\sigma_2$ and $\alpha$. We must count all configurations to determine the 
energy distribution function $f[E]$. That can be done in the following way:
\begin{eqnarray}
 f\left[E\right] \propto  &&\sum_{\sigma_1,\sigma_2}\int d\alpha dE_1dE_2 dL_1 dL_2 \bar{g}\left(E_1\right) \bar{g}\left(E_2\right) g \left(L_1 \right) g\left(L_2\right)  \nonumber \\
 && \times \ h\left(E_{cm}\left(E_1,E_2,L_1,L_2,\sigma_1,\sigma_2\right),\alpha-\bar{\alpha}\right) 
 \delta \left(E-E_3\left(E_1,E_2,L_1,L_2,\sigma_1,\sigma_2,\alpha\right)\right) \ .
\label{spectrum}
 \end{eqnarray}
Here, $\bar{g}\left(E_1\right)$
,~$\bar{g}\left(E_2\right)$,~$g \left(L_1 \right)$ and 
$g\left(L_2\right)$ stand for the distribution of energies and angular momenta of the colliding particles. 
$h\left(E_{cm}\left(E_1,E_2,L_1,L_2\right),\alpha-\bar{\alpha}\right)$ is the angular distribution of the massless particles in the center of mass frame. 
$\bar{\alpha}$ depicts the direction along which colliding particles
travel \eq{alphamas} which can be determined from $E_1$,~$E_2$
,~$L_1$ and $L_2$. 
The angular distribution is dictated by the differential cross section of the underlying particle physics process.
We assume that the angular distribution function $h$ takes the following form:
\begin{equation}
 h\left(E_{cm},\gamma\right)= \sum_{n=0}^{\infty}h_{n}\left(E_{cm}\right) \left(\cos\gamma\right)^n \ .
 \label{he}
\end{equation}
Functions $h_{n}\left(E_{cm}\right)$ capture the information about the physics at the energy $E_{cm}$.
To obtain the energy distribution $f(E)$, we integrate over all possible allowed values of conserved energies, angular momenta and angles subject to the constraint 
that their combination corresponds to the given energy $E$ which is enforced by the Dirac's delta function. In the end we normalize 
$f(E)$ over the entire range of the allowed energies.

\section{Ultra-high-energy collisions in the overspinning Kerr geometry}

In this section we describe the process of ultra-high-energy collisions in the overspinning Kerr geometry.
In the previous section we described the method we employ to study the collisional Penrose process in the Kerr spacetime. 
We now specialize to the overspinning Kerr spacetime geometry. We argued that for the divergence of the efficiency, 
it is necessary for particles to collide with divergent center-of-mass energy. We had identified the conditions under 
which it would be possible to have collisions with ultra-high-energy collisions in the overspinning Kerr geometry \cite{Patil1},\cite{Patil2}.
We recapitulate the results here to orient ourselves to analyze the collisional Penrose process.

\subsection{General discussion}
For the divergence of the center-of-mass energy of collision, it is necessary that $A$ must diverge. For $A$ to diverge,
$\Delta$ must take a value close to zero. 
From Eqs.~\eq{con1} and \eq{ecminfty},
\begin{equation}
 E_{cm}=m\sqrt{A^2-B^2-C^2} \rightarrow \infty ~~ \implies ~~ A \rightarrow \infty ~~ \implies ~~ \Delta \rightarrow 0 \ .
 \label{con2}
\end{equation}
For the overspinning Kerr geometry $\Delta=\left(r^2-2Mr+a^2\right)=\left(r-M\right)^2+\left(a^2-M^2\right)>0$ since $a>M$. 
The radial coordinate $r=r_{min}$ at which $\Delta$ is minimum and its minimum value $\Delta_{min}$ are given by 
\begin{equation}
 r_{min}=M ~~ ; ~~ \Delta_{min}=a^2-M^2 \ .
 \label{Dmin}
\end{equation}
Thus, we consider the collision at the radial location $r=r_{min}=M$ in the overspinning Kerr geometry with the spin parameter 
transcending the extremal value by an infinitesimal amount, i.e. ,
\begin{eqnarray}
r_{collision}&=&M \ ,\nonumber \\
a=M\left(1+\epsilon\right)~ &;& ~\epsilon \rightarrow 0^{+} \ .
\label{colra}
\end{eqnarray}
It is clear from Eq.~\eq{Dmin} that the
smaller the value of $|B|$ is, the larger
the center of mass energy of collision will be.
For the fixed values of $E_1$,$E_2$,$L_1$ and $L_2$; $|B|$ takes the smaller value if one of the particles is traveling in the 
radially inward direction and other particle is moving radially outwards, i.e., $\sigma_1\sigma_2=-1$, 
than the case where both the particles move in the radially inwards or radially outwards,
i.e., $\sigma_1\sigma_2=+1$. 
From Eqs.~\eq{ABC} and \eq{Dmin}, we can infer that 
\begin{eqnarray}
 &&|B(E_1,E_2,L_1,L_2,\sigma_1 \sigma_2=-1)| < |B(E_1,E_2,L_1,L_2,\sigma_1 \sigma_2=1)| \nonumber \\ 
 &&\implies  E_{cm}(E_1,E_2,L_1,L_2,\sigma_1 \sigma_2=-1) > E_{cm}(E_1,E_2,L_1,L_2,\sigma_1 \sigma_2=1) \ .
\label{eq35}
\end{eqnarray}
Therefore, we would like to consider the case where one of the particles travels in the radially outward direction and other 
particle travels in the radially inward direction. Without the loss of generality we assume that the 
particle $1$ moves in the radially outward direction. 
\begin{equation}
 \sigma_1=+1~;~\sigma_2=-1 
 \label{sigpm}
\end{equation}
We also assume that both the particles start from rest at infinity. In other words, particles are non-relativistic when they 
are faraway from the singularity, which is a reasonable assumption. When
particles fall towards singularity, they attain 
relativistic velocities in the high-curvature region. From Eq.~\eq{velmav}, we can set the conserved energies of two particles 
to unity, i.e.,
\begin{equation}
 E_1=E_2=1 \ .
 \label{E121}
\end{equation}

\begin{figure}
\begin{center}
\includegraphics[width=0.9\textwidth]{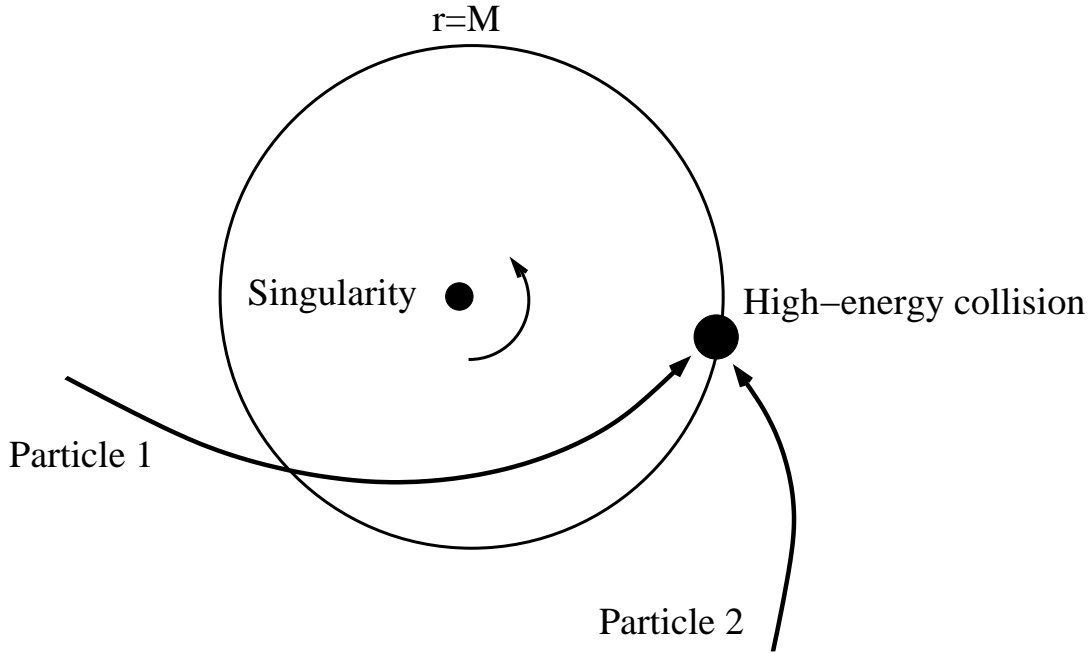}
\caption{ High-energy collision at $r=M$ around the Kerr naked singularity. Particle $1$ is an initially ingoing particle which 
turns back inside the circle $r=M$ and reappears at $r=M$ as an outgoing particle. Particle $2$ is an ingoing particle at $r=M$.
The center-of-mass energy of collision shows divergence when the Kerr spin parameter transcends the extremal value
by a small amount i.e. $a=M\left(1+\epsilon\right) \ ; \ \epsilon \rightarrow 0^{+}$.
}
\label{collision}
\end{center}
\end{figure}

\subsection{Allowed range for the angular momenta of the two particles}

Particle $1$ starting from infinity which is initially infalling must turn back at $r<M$ so as to appear at $r=M$
as an outgoing particle. Whereas particle $2$ should appear at $r=M$ as an ingoing particle. 
For this to happen the angular momenta of the two particles must be in the appropriate range.

Turning points for the particle with conserved energy $E=1$ and angular momentum per unit mass $L$ 
can be obtained from Eq.~\eq{velmav} by solving 
\begin{equation}
 \frac{2M}{r}-\frac{L^2}{r^2}+\frac{2M}{r^3}(L-a)^2=0
 \label{turp}
\end{equation}
and are given by
\begin{equation}
 r=r_{\pm}=\frac{L^2}{4M}\left(1\pm\sqrt{1-\frac{16M^2\left(L-a\right)^2}{L^4}}\right) \ .
 \label{rpm}
\end{equation}
The outer turning point $r=r_{+}$ is relevant for our discussion since we are interested in the particles which 
fall towards the singularity from infinity.
The turning points \eq{rpm} exist for the angular momenta in the following range: 
\begin{equation}
L<L_{-}=-2M \left(\sqrt{1+\frac{a}{M}}+1\right),~~
2M \left(\sqrt{1+\frac{a}{M}}-1\right)=L_{+}<L.
 \label{Lpm}
\end{equation}
As we increase the angular momentum above $L_{+}$, i.e., for $L>L{+}$ or decrease it below $L_{-}$, i.e.,for $L<L_{-}$, 
the outer turning point $r_{+}$ shifts radially outwards. The minimum values of the radial coordinates 
of the turning point, $r_{++}$ and $r_{+-}$,
for the positive and negative angular momenta, from Eqs.~\eq{rpm} and \eq{Lpm} are given by
\begin{equation}
 r_{++}=M\left(\sqrt{1+\frac{a}{M}}-1\right)^2~~\text{and}~~r_{+-}=M\left(\sqrt{1+\frac{a}{M}}+1\right)^2 \ ,
 \label{rpprmm}
\end{equation}
respectively. Note that $r_{+-}$ is always above the collision point, whereas $r_{++}$ is below the collision point for sufficiently 
small values of the spin parameter. That is, 
\begin{equation}
 M<a<3M \implies r_{++}<M ~~\text{and}~~ r_{+-}>M \ .
 \label{a3m}
\end{equation}
Since we are interested in the spin parameter which is slightly above the extremality, $r_{++}$ is below the collision point. 
The angular momentum required for the particle to turn back at $r=M$ obtained from Eq.~\eq{turp} is given by 
\begin{equation}
 L_{+,m}= 2a-\sqrt{2a^2-2M^2} \ .
 \label{lm}
\end{equation}

Particle $1$ must admit a turning point at the radial coordinate $r<M$ so that it appears at $r=M$ as 
an outgoing particle starting from infinity as an ingoing particle. It must turn back between $r=r_{++}$ and $r=M$.
Thus, from Eqs.~\eq{Lpm},~\eq{rpprmm} and \eq{lm}, its angular momentum $L_1$ must be in the following range:
\begin{equation}
2M \left(\sqrt{1+\frac{a}{M}}-1\right)<L_1< \left(2a-\sqrt{2a^2-2M^2}\right) \ .
 \label{L1}
\end{equation}

Particle $2$ appears at $r=M$ as an ingoing particle starting from infinity. Thus,~it should not admit a turning point at $r>M$. 
Thus,~from Eqs.~\eq{Lpm},~\eq{rpprmm} and \eq{lm}, its angular momentum $L_2$ must satisfy the condition
\begin{equation}
-2M\left(1+\sqrt{1+\frac{a}{M}}\right)<L_2< \left(2a-\sqrt{2a^2-2M^2}\right) \ .
 \label{L2}
\end{equation}

Thus{\bf ,~}we obtained the conditions that must be imposed on the angular momenta of the two colliding particles so that we have 
a collision in the desired setting. The angular momentum of particle $1$ must take a positive value, while the angular momentum 
of particle $2$ can take a positive as well as negative value.

\subsection{Center-of-mass energy of collision}

We now compute the center-of-mass energy of collision. The collision takes place at the radial coordinate $r=M$ and
is between particle $1$ that is radially outgoing and particle $2$ that is radially ingoing. 

We calculate $A$,~$B$ and $C$ to the leading order. From Eqs.~\eq{ABC},~\eq{colra},~\eq{sigpm},~\eq{E121},~\eq{L1} and \eq{L2},
we obtain 
\begin{eqnarray}
 A = \frac{\left(4M-\left(L_1 +L_2\right)\right)}{\sqrt{2 \epsilon}M} \ , 
B = \frac{\left(L_2-L_1\right)}{\sqrt{2 \epsilon}M} \ , 
C = \frac{\left(L_1 +L_2\right)}{2M} \ .
\label{abc} 
\end{eqnarray}
The center-of-mass energy of collision to the leading order obtained from Eqs.~\eq{con2} and \eq{abc} is given by 
\begin{equation}
 E_{cm}=m\sqrt{A^2-B^2-C^2}=\sqrt{\frac{2}{\epsilon}}\frac{m}{M}\sqrt{\left(2M-L_1\right)\left(2M-L_2\right)} \ .
 \label{ecmns}
\end{equation}
In the near-extremal limit, as $\epsilon \rightarrow 0$, the center-of-mass energy shows divergence
\begin{equation}
 \lim_{\epsilon\rightarrow 0^{+}} E_{cm} \rightarrow \infty \ .
 \label{ecmdiv}
\end{equation}

\subsection {Sign of B and C}
We now determine the sign of $B$ and $C$. The sign of $B$ which is the radial component of net velocity of the two 
colliding particles, depending on whether it is positive, negative or zero, determines whether the center of 
mass moves in the radially outward direction, radially inward direction 
or does not move in the radial direction. 

From Eqs.~\eq{ABC},~\eq{colra},~\eq{sigpm} and \eq{E121} we obtain 
\begin{equation}
 B=\frac{1}{\sqrt{a^2-M^2}}\left(\sqrt{2M^2-L_1^2+2\left(L_1-a\right)^2}-\sqrt{2M^2-L_2^2+2\left(L_2-a\right)^2}\right) \ .
 \label{BM}
\end{equation} 
It is clear from Eq.~\eq{BM} that depending on whether $\bar{B}$ is greater than, less than or equal to zero,
we can have $B$ positive, negative or zero, respectively, where 
\begin{equation}
 \bar{B}= \left(
   -L_1^2+2\left(L_1-a\right)^2\right)-\left(-L_2^2+2\left(L_2-a\right)^2\right)=\left(L_2-L_1\right)\left(4a-\left(L_1+L_2\right)\right) \ .
\end{equation}
From Eqs.~\eq{L1} and \eq{L2} we know that $\left(4a-\left(L_1+L_2\right)\right)>+2\sqrt{2a^2-2M^2}>0$. 
Therefore, the sign of $\bar{B}$ and thus of $B$ is determined by $\left(L_2-L_1\right)$.

From Eqs.~\eq{ABC},~\eq{Dmin} and \eq{E121}, we obtain 
\begin{equation}
C=\frac{\left(L_1+L_2\right)}{\sqrt{M^2+3a^2}}
\label{CM}
\end{equation}
Thus, the sign of $C$ is determined by $\left(L_1+L_2\right)$. 

We tabulate the different cases below. The values of $L_1$ and $L_2$ are assumed to be in their allowed range given by Eqs.~\eq{L1} and \eq{L2}.
\begin{table}[h!]
  \begin{center}
 \begin{tabular}{| p{3.5cm} | p{4.5cm} |}
  \hline
   $B$ and $C$ & $L_{1}$ and $L_{2}$ \\ \hline 
  \hline 
  $B>0~\text{and}~C>0$ & $L_2>L_1$ \\ \hline
  $B=0~\text{and}~C>0$ & $L_2=L_1$ \\ \hline
  $B<0~\text{and}~C>0$ & $L_2<L_1~\text{and}~L_2>-L_1 $ \\ \hline
  $B<0~\text{and}~C=0$ & $L_2<L_1~\text{and}~L_2=-L_1 $ \\ \hline
  $B<0~\text{and}~C>0$ & $L_2<L_1~\text{and}~L_2<-L_1 $ \\ \hline
  \end{tabular}
  \caption{Sign of $B$ and $C$. We tabulate the sign of $B$ and $C$, depending on the values of $L_1$ and $L_2$ in the allowed range.}
  \label{tab1}
\end{center}
\end{table}

While analyzing the collisional Penrose process, we must deal with the
cases $B>0$,~$B<0$ and $B=0$, separately. 
In case of the overspinning Kerr geometry, the results as we show would
be qualitatively the same irrespective of the sign of $B$,
which is not the case for Kerr black holes \cite{Berti}.

\section{Collisional Penrose process in the overspinning Kerr geometry}

In the previous section we described the process of ultra-high-energy collisions in the overspinning Kerr geometry 
which is the pre-requisite to the efficient energy extraction. In this section we analyze the collisional Penrose process 
and show that its efficiency shows divergence. We assume that two massless particles are produced in the collision. 
We obtain the escape cones for the massless particle to reach to infinity and energy of the particle within the escape cone. 

The collision point which is at $r=M$ is within $r=r_{m}\sim 4M$ as it can be shown from Eqs.~\eq{rmbm} and \eq{colra}. Thus,
the escape cones for the massless particles are given by Eq.~\eq{ecless}, i.e.,
\begin{equation}
 EC= \left[\left[\beta_1,\beta_2\right]\right]\cup \left[\left[\beta_3,\beta_4\right]\right] \ ,
\label{ECn}
\end{equation}
where the critical angles appearing in the expression above are given by
\begin{eqnarray}
\beta_1 \ =\ \alpha_1 &&= \ \alpha\left(\sigma=-1,b=a\right) \ ,\nonumber \\
\beta_2 \ =\ \alpha_2 &&= \ \alpha \left(\sigma=-1,b=2a-\sqrt{a^2-M^2}\right) \ ,\nonumber \\
\beta_3 \ =\ \alpha_5 &&= \ \alpha\left(\sigma=+1,b=2a-\sqrt{a^2-M^2}\right) \ ,\nonumber 
\end{eqnarray}
\begin{eqnarray}
\beta_4 \ =\ \alpha_7 &&= \ \alpha\left(\sigma=+1,b=a-\frac{\left(M^{\frac{2}{3}}+\left(a+\sqrt{a^2-M^2}\right)^{\frac{2}{3}}\right)^3}{\left(a+\sqrt{a^2-M^2}\right)}\right) \ .
\label{ECns}
\end{eqnarray}

\subsection{Center of mass moves radially outwards: $B>0$}

We now analyze the collisional Penrose process in the case where the center of mass moves
in the radially outward direction, i.e., $B>0$.
From Table \ref{tab1}, it happens when $L_2>L_1$ and consequently we also have $C>0$. 
The behavior of $A$,~$B$ and $C$ deduced from Eqs.~\eq{ABC},~\eq{colra},~\eq{sigpm} and \eq{E121} is as depicted below
\begin{equation}
 A = O\left(\epsilon^{-\frac{1}{2}}\right) \ , \  B = O\left(\epsilon^{-\frac{1}{2}}\right) \ , \  C = O\left(\epsilon^{0}\right) \ .
\label{abc>0}
 \end{equation}
The first colliding particle which is an outgoing particle at $r=M$ moves along the direction which makes angle $\alpha_{c,1}$, which 
can be obtained from Eqs.~\eq{alphamas},~\eq{colra},~\eq{sigpm} and \eq{E121}, is given by 
\begin{equation}
 \alpha_{c,1}=-\sqrt{\frac{\epsilon}{2}} \left[\left(\frac{L_1}{L_2-L_1}\right)\sqrt{\frac{2M-L_2}{2M-L_1}}+
 \left(\frac{L_2}{L_2-L_1}\right)\sqrt{\frac{2M-L_1}{2M-L_2}} \right] \  .
 \label{alphac1}
\end{equation}
In the near-extremal limit,  as $\epsilon \rightarrow 0^{+}$, we have $\alpha_{c,1} \rightarrow 0$. 
Thus, the particle moves almost along 
$+\hat{r}$ direction in the center-of-mass frame. Whereas, the second colliding particle 
moves almost along $-\hat{r}$ direction.

We now calculate the critical angles $\beta_1$,~$\beta_2$,~$\beta_3$ and $\beta_4$ from 
 Eqs.~\eq{alphamal},~\eq{colra},~\eq{sigpm},~\eq{E121} and \eq{ECns}, which is given by:
\begin{eqnarray}
 \beta_1 &=& \pi - \sqrt{2\epsilon}
  \frac{L_2}{\left(L_2-L_1\right)}\sqrt{\frac{2M-L_2}{2M-L_1}}  \ ,\nonumber \\
  \beta_2&=& \beta_3  = \pi-\arcsin \left(\frac{2\sqrt{2M-L_2}\sqrt{2M-L_1}}{\left(4M-L_1-L_2\right)}\right) \ , \nonumber \\
 \beta_4 &=&
  -\frac{\sqrt{2\epsilon}}{9}\sqrt{\frac{2M-L_1}{2M-L_2}}\frac{\left(L_1+8L_2\right)}{\left(L_2-L_1\right)} \ .
 \label{al123b>0}
\end{eqnarray}
In the near-extremal limit $\epsilon
\rightarrow 0^{+}$, the limit values of the critical angles are given by $\beta_1 \rightarrow \pi$ and 
$\beta_4 \rightarrow 0$.
$\beta_2=\beta_3$ since $r=M$ is a turning point for $b=2a-\sqrt{a^2-M^2}$ as we had shown earlier in Eq.~\eq{lm}.
A radially outgoing massless particle generated in the collision will escape to infinity if it is emitted along the angle which lies in the range 
$\alpha \in \left[\left[\beta_3,\beta_4\right]\right]$ and radially ingoing massless particle escapes to infinity if it is emitted along the angle
$\alpha \in \left[\left[\beta_1,\beta_2\right]\right]$.
\begin{figure}
\begin{center}
\includegraphics[width=0.7\textwidth]{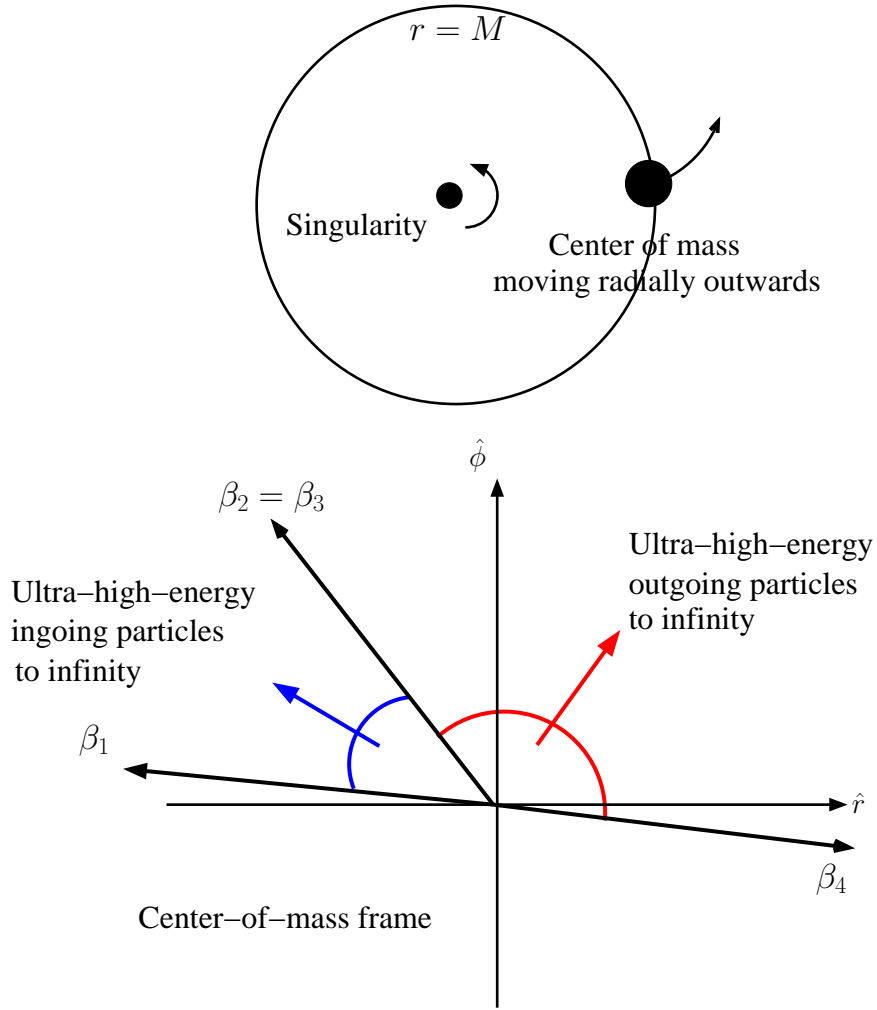}
\caption{ In the case $B>0$, the center of mass moves in the radially outward direction. Massless particles that are emitted in the
angular range $\left[\left[\beta_1,\beta_2\right]\right]$ move radially inwards, encounter a turning point and then escape to infinity. They are depicted 
by the blue arrow. Massless particles emitted in the angular range $\left[\left[\beta_3,\beta_4\right]\right]$ are emitted 
in the radially outward direction and escape to infinity. They are depicted by the red arrow. Particles reach infinity with 
ultra-high-energy unless $\alpha$ is close to $0$ or $\pi$.
}
\label{B>0}
\end{center}
\end{figure}
The escape fraction, i.e., the probability for the massless particle to escape to infinity assuming the isotropic distribution for the emission
in the center-of-mass frame is given by the ratio of the angular range of escape cones to the entire angular range. We can find
\begin{equation}
 E.F.= \frac{|\beta_1-\beta_2|+|\beta_3-\beta_4|}{2\pi} \rightarrow
  \frac{1}{2} \  , 
 \label{efb>0}
\end{equation}
in the limit $\epsilon \rightarrow 0^{+}$.
Therefore, in the near-extremal limit,
the escape fraction is half.
Thus, if we consider a large number of collisions, then 
half of the particles produced in the collisions will escape to infinity.

The energy of the massless particle as measured by an observer at infinity is obtained from 
Eqs.~\eq{E},~\eq{colra},~\eq{sigpm} and \eq{E121}.
Unless the particle is emitted along the angle which is
very close to $0$ or $\pi$, to the leading order it is given by
\begin{equation}
  E(\alpha)=\frac{m}{\sqrt{2}M}\frac{\sqrt{2M-L_2}\sqrt{2M-L_1}}{\sqrt{\epsilon}} \sin \alpha \ .
  \label{EB>0}
\end{equation}
The energy shows divergence in the near-extremal limit, i.e.,
\begin{equation}
 \lim_{\epsilon \rightarrow 0^{+}} E \rightarrow \infty \ .
 \label{Einf1}
\end{equation}
The energy of the other particle which is produced in the collision and moves in the opposite direction is given by
\begin{equation}
 E'(\alpha)=-\frac{m}{\sqrt{2}M}\frac{\sqrt{2M-L_2}\sqrt{2M-L_1}}{\sqrt{\epsilon}} \sin \alpha \ .
  \label{eq96a}
\end{equation}
It takes a large negative value, i.e.,
\begin{equation}
 \lim_{\epsilon \rightarrow 0^{+}} E' \rightarrow -\infty \ .
  \label{Epinf1}
\end{equation}
Thus,~the collisional Penrose process is at work and is responsible for 
ultra-high-energy particle escaping to infinity.

If the massless particle emitted along the angle which is sufficiently close to $0$ or $\pi$, the energy turns out to be 
finite. For instance for the massless particle traveling along the angles $\beta_1$ and $\beta_4$, their energies obtained 
from Eqs.~\eq{E},~\eq{colra},~\eq{sigpm} and \eq{E121} are given by
\begin{eqnarray}
E\left(\beta_1\right) &=& \frac{m}{M}\left(2M-L_2\right) \ 
\end{eqnarray}
and
\begin{eqnarray}
E\left(\beta_4\right) &=& \frac{1}{9}\frac{m}{M}\left(2M-L_1\right) \ ,
\label{e3fin1}
\end{eqnarray}
respectively.

However the proportion of the particles which escape to infinity with finite energy is minuscule. Almost all the particles
that escape to infinity are ultra-high-energy particles with divergent energies.

\subsection{Center of mass does not move in the radial direction: $B=0$}

Here we consider the case where the center of mass does not move in the radial direction, i.e., $B=0$.
From Table \ref{tab1}, it happens when $L_2=L_1=L$ and we have $C>0$.
The behavior of $A$,~$B$ and $C$ can be deduced from Eqs.~\eq{ABC},~\eq{colra},~\eq{sigpm} and \eq{E121} and is given by
\begin{equation}
 A = O\left(\epsilon^{-\frac{1}{2}}\right) \ , \  B=0 \ , \  C = O\left(\epsilon^{0}\right) \ .
 \label{abc=0}
 \end{equation}
The first colliding particle moves along the direction which makes angle $\alpha_{c,1}$, which 
can be obtained from Eqs.~\eq{alphamas},~\eq{colra},~\eq{sigpm} and \eq{E121}, is given by 
\begin{equation}
 \alpha_{c,1}=-\frac{\pi}{2} \ .
 \label{alphac12}
\end{equation}
First particle moves almost along 
$-\hat{\phi}$ direction in the center-of-mass frame. The second colliding particle 
moves almost along $+\hat{\phi}$ direction.

We now calculate the critical angles $\beta_1$,~$\beta_2$,~$\beta_3$ and $\beta_4$ from Eqs.~
\eq{alphamal},~\eq{colra},~\eq{sigpm},~\eq{E121} and \eq{ECns} as follows:
\begin{eqnarray}
 \beta_1 &=&  \frac{\pi}{2}-\frac{\sqrt{2\epsilon}M}{2M-L} \ , \nonumber \\
 \beta_2&=&\beta_3 = 0 \ , \nonumber \\
 \beta_4 &=&
  -\frac{\pi}{2}-\frac{\sqrt{2\epsilon}\left(7M+L\right)}{9\left(2M-L\right)}
  \ ,
 \label{al123b=0}
\end{eqnarray}
as $\epsilon\to 0^{+}$.
The limit values of the critical angles in the near-extremal limit are given by $\beta_1 \rightarrow \frac{\pi}{2}$ and
$\beta_4 \rightarrow -\frac{\pi}{2}$.
Again $\beta_2=\beta_3$ since $r=M$ is a turning point for
$b=2a-\sqrt{a^2-M^2}$ as we had shown earlier in Eq.~\eq{lm}.
A radially outgoing massless particle generated in the collision will escape to infinity if it is emitted along an angle which lies in the range 
$\alpha \in \left[\left[\beta_3,\beta_4\right]\right]$ and a radially ingoing massless particle escapes to infinity if it is emitted along an angle
$\alpha \in \left[\left[\beta_1,\beta_2\right]\right]$. 

\begin{figure}
\begin{center}
\includegraphics[width=0.7\textwidth]{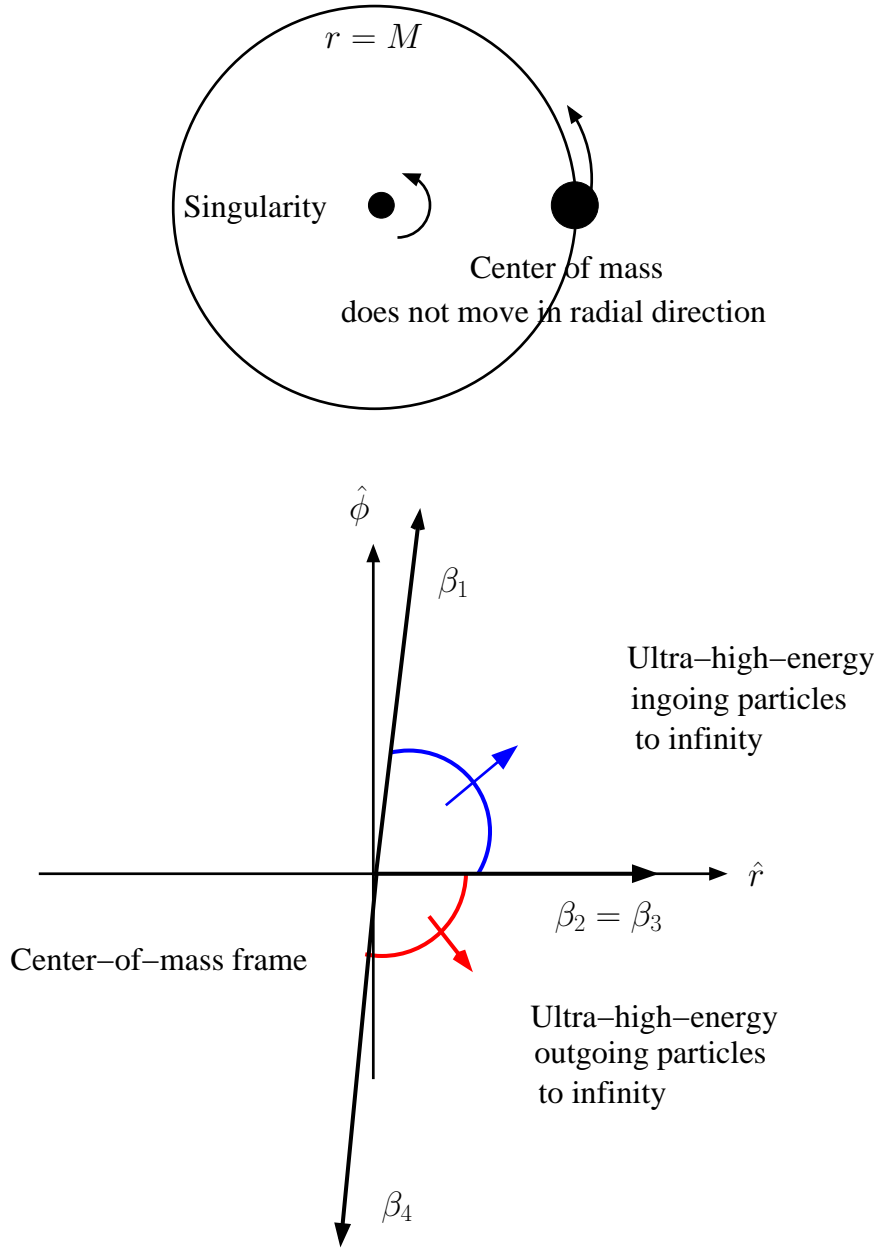}
\caption{ In the case $B=0$, the center of mass does not move in the radial direction. 
Massless particles that are emitted in the
angular range $\left[\left[\beta_1,\beta_2\right]\right]$ move radially inwards, encounter a turning point and then escape to infinity. 
They are depicted by the blue arrow. Massless particles emitted in the angular range $\left[\left[\beta_3,\beta_4\right]\right]$ are emitted 
in the radially outward direction and escape to infinity. They are depicted by the red arrow. Particles reach infinity with 
ultra-high-energy unless $\alpha$ is close to
 $-{\pi}/{2}$ or ${\pi}/{2}$.
}
\label{B=0}
\end{center}
\end{figure}

The escape fraction assuming the isotropic distribution for the emission
in the center-of-mass frame is given by 
\begin{equation}
 E.F.= \frac{|\beta_1-\beta_2|+|\beta_3-\beta_4|}{2\pi} \rightarrow
  \frac{1}{2} \  ,
 \label{efb=0}
\end{equation}
in the limit $\epsilon\to 0^{+}$. Therefore, in
the near-extremal limit, the escape fraction is half.

The energy of the massless particle as measured by an observer at infinity is obtained 
from Eqs.~\eq{E},~\eq{colra},~\eq{sigpm} and \eq{E121}.
For the particles emitted along the angles not 
very close to $0$ or $\pi$, energy to the leading order is given by
\begin{equation}
  E(\alpha)=\frac{m}{\sqrt{2}M}\frac{\sqrt{2M-L_2}\sqrt{2M-L_1}}{\sqrt{\epsilon}} \cos \alpha \ .
  \label{EB=0}
\end{equation}
Thus{\bf ,~}the energy shows divergence in the near-extremal limit, i.e.,
\begin{equation}
 \lim_{\epsilon \rightarrow 0^{+}} E \rightarrow \infty \ .
 \label{Einf2}
\end{equation}
The energy of the other particle moving in the opposite direction is given by
\begin{equation}
 E'(\alpha)=-\frac{m}{\sqrt{2}M}\frac{\sqrt{2M-L_2}\sqrt{2M-L_1}}{\sqrt{\epsilon}} \cos \alpha \ .
  \label{eq96b}
\end{equation}
It tends to negative infinity, i.e.,
\begin{equation}
 \lim_{\epsilon \rightarrow 0^{+}} E' \rightarrow -\infty .
  \label{Epinf1}
\end{equation}
This confirms the fact that the collisional Penrose process is at work and is responsible for 
the generation of ultra-high-energy particle escaping to infinity.

If the massless particle emitted along the angle which is sufficiently close to $0$ or $\pi$, the energy turns out to be 
finite. For instance for the massless particle traveling along angles $\beta_1$ and $\beta_4$, the energies obtained 
from Eqs.~\eq{E},~\eq{colra},~\eq{sigpm} and \eq{E121} are given by
\begin{eqnarray}
E\left(\beta_1\right) &=& \frac{m}{M}\left(2M-L\right) \ ,
\end{eqnarray}
and 
\begin{eqnarray}
E\left(\beta_4\right) &=& \frac{1}{9}\frac{m}{M}\left(2M-L\right) \ ,
\label{e3fin2}
\end{eqnarray}
respectively.

The proportion of the particles which escape to infinity with finite energy is minuscule. Almost all the particles
that escape to infinity are ultra-high-energy particles.

\subsection{Center of mass moves radially inwards: $B<0$}

We now consider the case where the center of mass moves in the direction radially inwards, i.e., $B<0$.
From Table \ref{tab1}, it happens when $L_2<L_1$ and we can have $C$ either positive, negative or zero.
The behavior of $A$ ,~$B$ and $C$,~ as inferred from Eqs.~\eq{ABC},~\eq{colra},~\eq{sigpm} and \eq{E121}, is given by
\begin{equation}
 A = O\left(\epsilon^{-\frac{1}{2}}\right) \ , \  B = O\left(\epsilon^{-\frac{1}{2}}\right) \ , \  C = O\left(\epsilon^{0}\right) \ .
 \label{abc<0}
 \end{equation}
The angle $\alpha_{c,1}$, along which the first colliding particle moves,
can be obtained from Eqs.~\eq{alphamas},~\eq{colra}
,~\eq{sigpm} and \eq{E121}, and is given by 
\begin{equation}
 \alpha_{c,1}=\pi +\sqrt{\frac{\epsilon}{2}}\left(\left(\frac{L_1}{L_1-L_2}\right)\sqrt{\frac{2M-L_2}{2M-L_1}}+
 \left(\frac{L_2}{L_1-L_2}\right)\sqrt{\frac{2M-L_1}{2M-L_2}}\right) \ ,
 \label{alphac13}
\end{equation}
in the limit $\epsilon \rightarrow 0^{+}$. The first particle moves almost along 
$-\hat{r}$ direction, whereas the second colliding particle 
moves almost along $+\hat{r}$ direction in the center-of-mass frame.

We now calculate the critical angles $\beta_1$,~$\beta_2$,~$\beta_3$ and $\beta_4$ from Eqs.~
\eq{alphamal},~\eq{colra},~\eq{sigpm},~\eq{E121} and \eq{ECns} as follows:
\begin{eqnarray}
 \beta_1 &=&  \sqrt{2\epsilon} \frac{L_2}{\left(L_1-L_2\right)}\sqrt{\frac{2M-L_2}{2M-L_1}} \ , \nonumber \\
 \beta_2 &=& \beta_3 = -\pi+\arcsin \left(\frac{2\sqrt{2M-L_2}\sqrt{2M-L_1}}{\left(4M-L_1-L_2\right)}\right) \ , \nonumber \\
\beta_4 &=& -\pi +\frac{\sqrt{2\epsilon}}{9}\sqrt{\frac{2M-L_1}{2M-L_2}}\frac{\left(L_1+8L_2\right)}{\left(L_1-L_2\right)} \ .
\label{al123b<0}
\end{eqnarray}
The limit values of critical angles in the near-extremal limit $\epsilon \rightarrow 0^{+}$
are given by $\beta_1 \rightarrow 0$ and 
$\beta_4 \rightarrow -\pi$. We have $\beta_2=\beta_3$ from Eq.~\eq{lm}.
The massless particle produced in the collision will escape to infinity if it is emitted 
along an angle which lies in the range 
$\alpha \in \left[\left[\beta_3,\beta_4\right]\right]$ and the radially ingoing massless particle escapes to infinity if it is emitted along angle
$\alpha \in \left[\left[\beta_1,\beta_2\right]\right]$. 
\begin{figure}
\begin{center}
\includegraphics[width=0.7\textwidth]{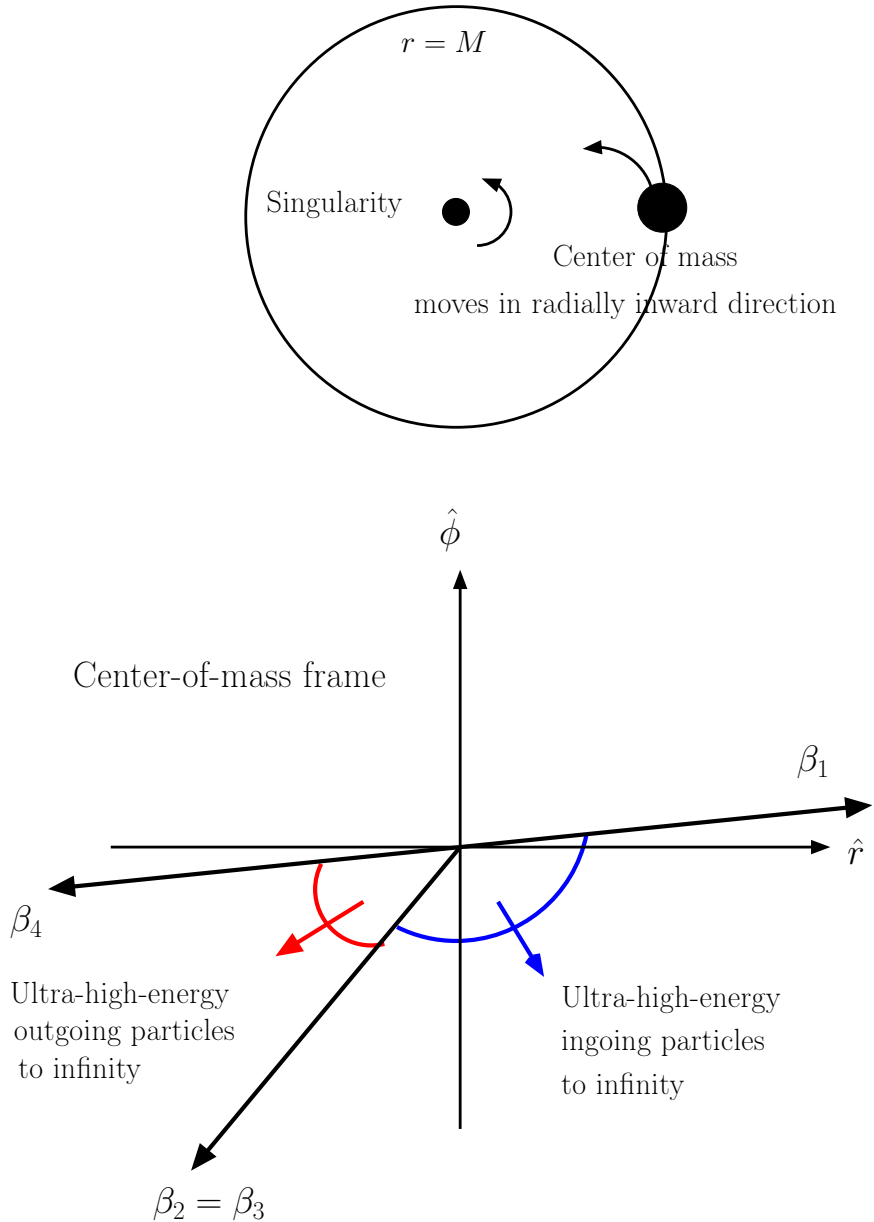}
\caption{ In the case $B<0$, the center of mass moves in the radially inward direction. 
Massless particles that are emitted in the
angular range $\left[\left[\beta_1,\beta_2\right]\right]$ move radially inwards, encounter a turning point and escape to infinity. 
They are depicted by the blue arrow. Massless particles emitted in the angular range $\left[\left[\beta_3,\beta_4\right]\right]$ are emitted 
in the radially outward direction and escape to infinity. They are depicted by the red arrow. Particles reach infinity with 
ultra-high-energy unless $\alpha$ is close to $0$ or $\pi$.
}
\label{B<0}
\end{center}
\end{figure}
The escape fraction if we assume isotropic distribution for the emission of massless particles
in the center-of-mass frame is given by
\begin{equation}
 E.F.= \frac{|\beta_1-\beta_2|+|\beta_3-\beta_4|}{2\pi} \rightarrow \frac{1}{2} \ .
 \label{efb<0}
\end{equation}
In the near-extremal limit, the escape fraction is again half.

The energy of the massless particle as measured by observer at infinity is obtained from 
Eqs.~\eq{E},~\eq{colra},~\eq{sigpm} and \eq{E121}.
Unless the angle $\alpha$ is 
very close to $0$ or $\pi$, its energy to the leading order is given by
\begin{equation}
  E(\alpha)=-\frac{m}{\sqrt{2}M}\frac{\sqrt{2M-L_2}\sqrt{2M-L_1}}{\sqrt{\epsilon}} \sin \alpha \ .
  \label{EB<0}
\end{equation}
Their energy shows divergence in the near-extremal limit
\begin{equation}
 \lim_{\epsilon \rightarrow 0^{+}} E \rightarrow \infty \ .
 \label{Einf3}
\end{equation}
The energy of the second particle moving in the opposite direction is given by
\begin{equation}
 E'(\alpha)=+\frac{m}{\sqrt{2}M}\frac{\sqrt{2M-L_2}\sqrt{2M-L_1}}{\sqrt{\epsilon}} \sin \alpha \ .
  \label{eq96c}
\end{equation}
It tends to negative infinity, i.e.,
\begin{equation}
 \lim_{\epsilon \rightarrow 0^{+}} E' \rightarrow -\infty \ .
  \label{Epinf3}
\end{equation}
Thus,~the collisional Penrose process is at work and accounts for the
the generation of ultra-high-energy particle escaping to infinity.

If the massless particle emitted along the angle which is sufficiently close to $0$ or $\pi$, its energy would be
finite. For instance, for the massless particle traveling along angles $\beta_1$ and $\beta_4$, their energies obtained 
from Eqs.~\eq{E},~\eq{colra},~\eq{sigpm}
and \eq{E121} are given by
\begin{eqnarray}
E\left(\beta_1\right) &=& \frac{m}{M}\left(2M-L_2\right) \ ,
\end{eqnarray}
and
\begin{eqnarray}
E\left(\beta_4\right) &=& \frac{1}{9}\frac{m}{M}\left(2M-L_1\right) \ ,
\label{e3fin3}
\end{eqnarray}
respectively.

The proportion of the particles which escape to infinity with finite energy is minuscule. Almost all the particles
that escape to infinity are ultra-high-energy particles.

\subsection{Divergence of efficiency of collisional Penorse process}

We now compute the efficiency of the collisional Penrose process. For further analysis, we find it convenient to 
rotate the coordinate axes by the angle ${\pi}/{2}$ in the case where $B=0$ 
and by the angle $\pi$ in the case $B<0$. From Eqs.~\eq{EB>0},~\eq{EB=0} and \eq{EB<0}, the 
energy of the particle that escapes to infinity can now be written as 
\begin{equation}
 E=\frac{m}{\sqrt{2}M} \frac{\sqrt{2M-L_2}\sqrt{2M-L_1}}{\sqrt{\epsilon}} \sin \alpha \ .
 \label{EU}
\end{equation}
The efficiency of the collisional Penrose process which is defined by the ratio of conserved energy of the particle 
escaping to infinity to the total conserved energy of the colliding particles is given by 
\begin{equation}
 \eta=\frac{E}{2m}=\frac{1}{2\sqrt{2}M} \frac{\sqrt{2M-L_2}\sqrt{2M-L_1}}{\sqrt{\epsilon}} \sin \alpha \ .
 \label{eff}
\end{equation}
For a fixed value of the parameter $\epsilon$ depicting the deviation of Kerr geometry from extremality, the
efficiency of collisional Penrose process is maximum when $\alpha=\frac{\pi}{2}$, i.e., the particle is emitted along the direction orthogonal to the 
direction in which the colliding particles travel and angular momenta $L_1$ and $L_2$ take minimum possible values in the 
allowed range \eq{L1} and \eq{L2}, namely $L_1=2M \left(\sqrt{1+\frac{a}{M}}-1\right)$ and $L_2=-2M \left(\sqrt{1+\frac{a}{M}}+1\right)$.
From Eq.~\eq{eff}, it is given by
\begin{equation}
 \eta_{max}=\frac{1}{\sqrt{\epsilon}} \ .
 \label{effmax}
\end{equation}
The maximum efficiency goes to infinity in the near-extremal limit where the spin parameter transcends the extremal value by an 
infinitesimally small number, i.e.,
\begin{equation}
\lim_{\epsilon \rightarrow 0^{+}} \eta_{max} \rightarrow \infty \ .
 \label{effdiv}
\end{equation}
Thus, we have demonstrated that the efficiency of collisional Penrose process can be arbitrarily large in the overspinning Kerr spacetime 
geometry. This implies that it will be possible to extract large energy and generate ultra-high-energy particles starting 
with the particles with moderate energies. In the next section, we explore the implications of the super-efficient 
collisional Penrose process from the point of view of astrophysics as well as fundamental particle physics.

\section{Bright spot of ultra-high-energy particles}

The angular momentum of the massless particles escaping to infinity obtained from Eqs.~\eq{L3},~\eq{colra},~\eq{sigpm} and \eq{E121}
given by the following expression to the leading order:
\begin{equation}
 L=\sqrt{2}m \frac{\sqrt{2M-L_2}\sqrt{2M-L_1}}{\sqrt{\epsilon}} \sin \alpha \ .
 \label{Lp}
\end{equation}
This expression is valid for all the three cases $B>0$, $B=0$ and $B<0$
after rotating the coordinate axes by the angles ${\pi}/{2}$ and $\pi$
in the last two cases. Here $\alpha$ is assumed to be away from $0$ and $\pi$ and thus it corresponds to the massless particles with ultra-high-energies. 
Note that the angular momentum is always positive for the particles that are emitted with large energies. Particles with positive 
angular momentum travel in the same way in which the naked singularity rotates. Thus, to the distant observer, 
they seem to emerge from that side of the singularity, which is rotating towards the observer. Further, the 
impact parameter of all ultra-high-energy particles, obtained from Eqs.~\eq{EU} and \eq{Lp}
is given by the approximate constant value given below:
\begin{equation}
b=\frac{L}{E}=2M \ .
 \label{bU}
\end{equation}
Thus, ultra-high-energy particles seem to emerge from a bright spot of a narrow width. 

\begin{figure}
\begin{center}
\includegraphics[width=0.7\textwidth]{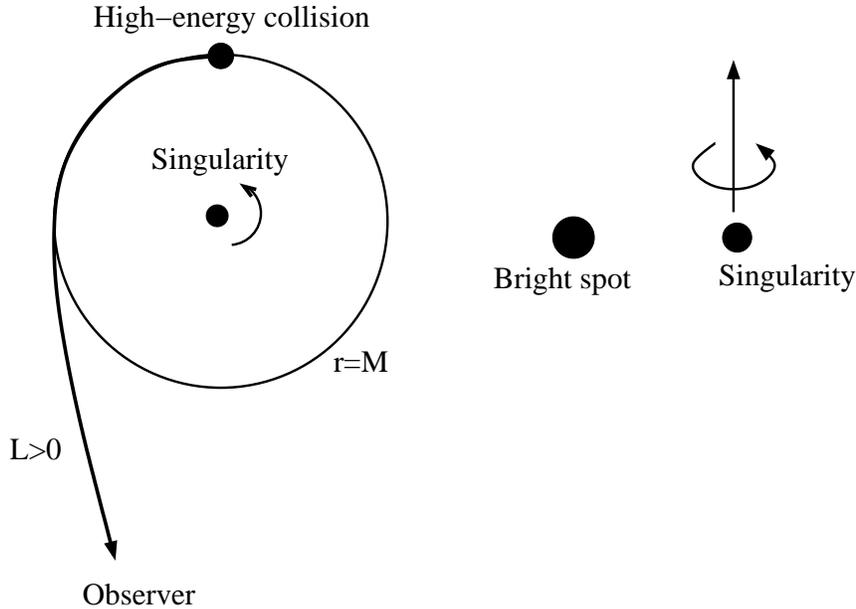}
\caption{ Bright spot. Left Panel: Ultra-high-energy 
particles created in the high-energy collision have a positive angular momentum. Thus, they co-rotate with the singularity. 
Right Panel: High-energy particles seem to originate from the bright spot on that side of the singularity which is rotating towards the observer.
}
\label{spot2}
\end{center}
\end{figure}

We now compute the angular location of the bright spot relative to the singularity as observed by a distant observer in
LNRF at the large value of the radial coordinate $r$. The results would not change much for the non-corotating observers since the 
frame-dragging is minuscule at large radii. Consider a radially outgoing massless particle with impact parameter $b$. 
The spatial components of velocity in LNRF can be read off from Eq.~\eq{lnrfmal}
\begin{eqnarray}
 U^{(\phi)}_{LNRF}(b,r) &=& \frac{b}{\sqrt{\left(r^2+a^2+\frac{2Ma^2}{r}\right)}} \ , \nonumber \\ \nonumber \\
 U^{(r)}_{LNRF}(b,r) &=& \frac{r}{\sqrt{\Delta}}\sqrt{1-\frac{\left(b^2-a^2\right)}{r^2}+\frac{2M\left(b-a\right)^2}{r^3}} \ , \nonumber \\ \nonumber \\
 U^{(\theta)}_{LNRF}(b,r) &=& 0  \ .
 \label{lnrfnzmal}
\end{eqnarray}
The three-velocity of the massless particle obtained from Eq.~\eq{lnrfnzmal} is given by
\begin{equation}
 V^{(i)}_{LNRF}(b,r)=\frac{U^{(i)}_{LNRF}(b,r)}{\sqrt{\delta_{jk}U^{(j)}_{LNRF}(b,r)U^{(k)}_{LNRF}(b,r)}} \ .
 \label{3velmal}
\end{equation}

The direction in which the particle travels as seen by LNRF observer is dictated by its three-velocity. 
Since all the ultra-high-energy massless particles produced in high-energy collisions around the naked singularity
have the impact parameter $b=2M$, the direction in which they travel is dictated by $V^{(i)}_{LNRF}(2M,r)$.
Consider a hypothetical massless particle with the zero impact parameter $b=0$ which emerges from the singularity and reaches
the distant observer. The trajectory of this particle is bent due to the frame-dragging in the Kerr spacetime. The direction 
in which it travels with respect to the observer is dictated by $V^{(i)}_{LNRF}(0,r)$.
The angular location of the bright spot from where high-energy particles appear to emerge from relative to
the singularity can be
obtained by computing the angle between $V^{(i)}_{LNRF}(2M,r)$ and $V^{(i)}_{LNRF}(0,r)$. It is given by 
\begin{equation}
 \sin \xi =\sqrt{1-\left(\delta_{jk}V^{(j)}_{LNRF}(2M,r)V^{(k)}_{LNRF}(0,r)\right)^2} \ .
 \label{sinxi}
\end{equation}
From Eqs.~\eq{lnrfnzmal},~\eq{3velmal} and \eq{sinxi}, we obtain
\begin{equation}
 \sin \xi= \frac{U^{(\phi)}_{LNRF}(2M,r)}{\sqrt{U^{(\phi) \ 2}_{LNRF}(2M,r)+U^{(r) \ 2}_{LNRF}(2M,r)}} \ .
 \label{sinxi2}
\end{equation}
Thus, in the large $r$ limit,
the expression for the angle $\xi$ obtained from Eqs.~\eq{lnrfnzmal} and \eq{sinxi2} is given by
\begin{equation}
 \xi = \frac{2M}{r} \ .
 \label{xi}
\end{equation}
This is the angular location of the bright spot on that side of the singularity, which is rotating towards us
from where high-energy particles appear to originate to the distant observer.

\section{Distribution of energy of massless particles}

We now compute the energy distribution of the ultra-high-energy massless particles escaping to infinity. 
As described earlier, we focus only on the collisions that occur at $r=M$ and compute their 
contribution to the spectrum. The collisions are between ingoing and outgoing particles, i.e., $\sigma_1=+1$ and $\sigma_2=-1$.
 The conserved energies of the colliding particles are taken to be unity $E_1=E_2=1$. This allows us to omit summation over 
$\sigma_1$ and $\sigma_2$ and also the integration over $E_1$ and $E_2$ in Eq.~\eq{spectrum}.
Thus, from Eqs.~\eq{spectrum},~\eq{ecmns} and \eq{EU}, we obtain the following expression
for the spectrum of the massless particles:
\begin{eqnarray}
 f\left[E\right] \propto \int && d\alpha dL_1 dL_2   g \left(L_1 \right) g\left(L_2\right)  \ h\left(\sqrt{\frac{2}{\epsilon}}\frac{m}{M}\sqrt{\left(2M-L_1\right)\left(2M-L_2\right)},\alpha\right) \nonumber \\
&& \times \delta \left(E-\frac{m}{\sqrt{2}M} \frac{\sqrt{2M-L_2}\sqrt{2M-L_1}}{\sqrt{\epsilon}} \sin \alpha \right) \ .
\label{spectrum1}
 \end{eqnarray}
We need to integrate over all possible values of 
angular momenta $L_1$ and $L_2$ and angles of emission $\alpha$ of massless particles in the center-of-mass frame. 

We integrate over angle $\alpha$ using the following property of Dirac's delta function:
\begin{equation}
\delta \left(f(x)\right)= \sum_{i}
 \frac{\delta\left(x-x_{i}\right)}{|f'(x_i)|} \ \text{,~where} \ f(x_i)=0 \ .
 \label{dd}
\end{equation}
From Eqs.~\eq{spectrum1} and \eq{dd},
\begin{eqnarray}
 f\left[E\right] && \propto \int  d\alpha dL_1 dL_2 g \left(L_1 \right) g\left(L_2\right) \nonumber \\
 && \frac{\left( h\left(\sqrt{\frac{2}{\epsilon}}\frac{m}{M}\sqrt{\left(2M-L_1\right)\left(2M-L_2\right)},\beta\right)  + h\left(\sqrt{\frac{2}{\epsilon}}\frac{m}{M}\sqrt{\left(2M-L_1\right)\left(2M-L_2\right)},\pi -\beta\right)  \right)}
 {\sqrt{\left(\frac{m}{\sqrt{2}M} \frac{\sqrt{2M-L_2}\sqrt{2M-L_1}}{\sqrt{\epsilon}}\right)^2 -E^2}} \ ,
 \label{sp3}
 \end{eqnarray}
where $\beta$ is an angle such that 
\begin{equation}
 \sin \beta=\frac{2E}{\sqrt{\frac{2}{\epsilon}}\frac{m}{M}\sqrt{\left(2M-L_1\right)\left(2M-L_2\right)}} \ .
 \label{eq104}
\end{equation}
We now use the expansion of angular distribution function $h$ we proposed earlier in Eq.~\eq{he}.
For the analysis carried out in this paper we retain only first two relevant terms. We obtain
\begin{eqnarray}
 f\left[E\right] \propto &&\int dL_1 dL_2  g \left(L_1 \right) g\left(L_2\right)
 \frac{h_0}{\sqrt{y\left(2M-L_1\right)\left(2M-L_2\right)-E^2}} \nonumber \\
+ &&\int dL_1 dL_2  g \left(L_1 \right) g\left(L_2\right) 
\frac{h_2\sqrt{y\left(2M-L_1\right)\left(2M-L_2\right)-E^2} }{\left(y\left(2M-L_1\right)\left(2M-L_2\right)\right)} \ ,
\label{sp4}
 \end{eqnarray}
where $ y={m^2}/({2M^2}{\epsilon})$. The first term in the expression above corresponds to the contribution 
from the isotropic emission, whereas the second term corresponds to the leading order contribution from anisotropic emission
in the center-of-mass frame. We also assume that neither $h_0$ nor $h_2$ varies significantly at higher energies and both 
can be taken as constant. This assumption
allows us to analytically calculate the spectrum. In a more realistic
calculation, the variation of $h_0$ and $h_2$
must be taken into account. $h_0$ and $h_2$ can be calculated from the differential 
cross section of underlying particle physics process. 

We now compute the energy distribution function by making different assumptions regarding the distribution of angular momenta.

\subsection{Both $L_1$ and $L_2$ are fixed}

\begin{figure}
\begin{center}
\includegraphics[width=0.9\textwidth]{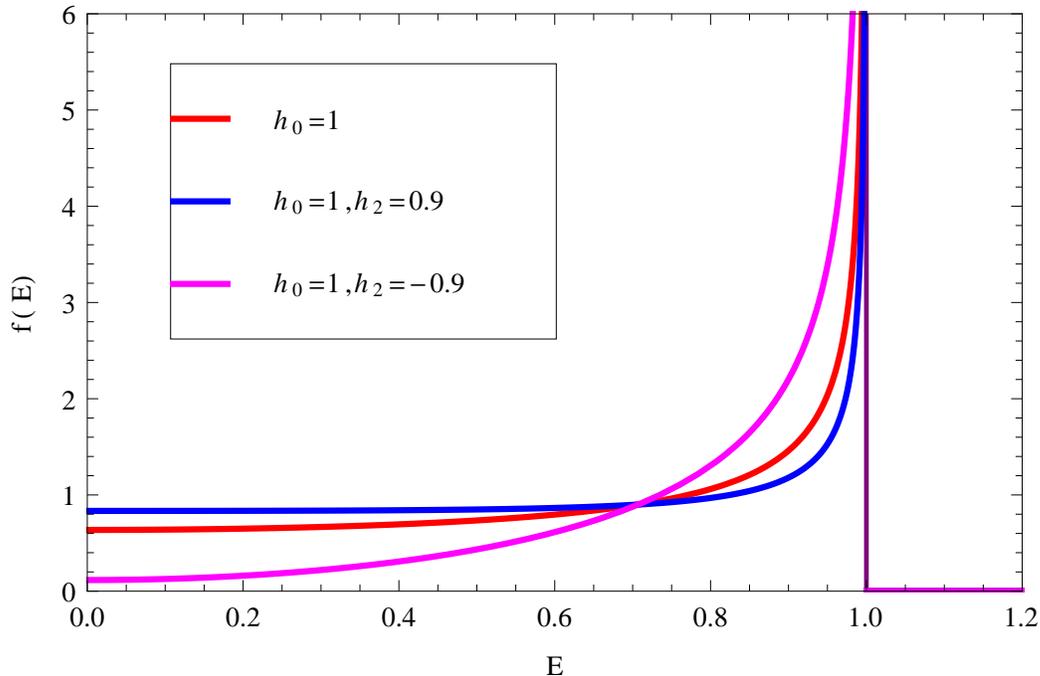}
\caption{ Energy distribution function $f[E]$ is plotted here, where $E$  
is expressed in units of $E_{m}$. We assume that the angular momenta of both the 
colliding particles take the fixed values. The distribution function shows divergence at $E=E_m$.
The plots are made for three different cases A. $h_0=1$,~$h_2=0$, B. $h_{0}=1$,~$h_2=0.9$, and  C. $h_0=1$,~$h_2=-0.9$.
}
\label{dist1}
\end{center}
\end{figure}

We first make a naive assumption that the angular momentum of particle $1$ takes a fixed value $L_1^{*}$ and 
that of particle $2$ takes a fixed value $L_2^{*}$ in the allowed range, i.e., the distribution 
of the angular momenta is  Dirac's delta function 
\begin{equation}
g \left(L_1 \right)=\delta \left(L_1-L_1^{*}\right) \ ; \ g \left(L_2 \right)=\delta \left(L_2-L_2^{*}\right) \ .
 \label{h12dd}
\end{equation}
From Eq.~\eq{h12dd}, we can integrate over $L_1$ and $L_2$. 
The energy distribution function, which is given by
Eq.~\eq{sp4}, can now be written as 
\begin{equation}
 f[E] \propto \left(\frac{h_0}{\sqrt{y\left(2M-L_1^{*}\right)\left(2M-L_2^{*}\right)-E^2}}
 +\frac{h_2\sqrt{y\left(2M-L_1^{*}\right)\left(2M-L_2^{*}\right)-E^2} }{\left(y\left(2M-L_1^{*}\right)\left(2M-L_2^{*}\right)\right)}\right) \ .
 \label{sp12dd}
\end{equation}
It is clear from the expression above that the highest allowed value of the energy $E$ is 
\begin{equation}
 E_{m}=\sqrt{y\left(2M-L_1^{*}\right)\left(2M-L_2^{*}\right)} \ .
 \label{em1}
\end{equation}
Thus,~the spectrum admits an upper cut-off at energy $E_m$ which can go 
to infinity in the near-extremal limit.
We normalize the distribution function \eq{sp12dd} so that 
\begin{equation}
 \int_{0}^{E_{m}} dE \ f(E)=1 \ .
 \label{nor1}
\end{equation}
From Eqs.~\eq{sp12dd} and \eq{nor1}, we
obtain a normalized distribution function as 
\begin{equation}
f[E] = \frac{2}{\pi\left(h_0+\frac{h_2}{2}\right)}\left(\frac{h_0}{\sqrt{E_m^2-E^2}}
 +\frac{h_2\sqrt{E_m^2-E^2} }{E_m^2}\right) \ .
 \label{nd1}
\end{equation}
 We plot $f[E]$ in Fig.~\ref{dist1}.
Quite remarkably, the distribution function shows divergence at $E=E_m$. This is an artifact of the choice
of Dirac's delta distribution for both the angular momenta and disappears when we relax this assumption as we will show later.

\subsection{$L_1$ is fixed and $L_2$ follows uniform distribution}

\begin{figure}
\begin{center}
\includegraphics[width=0.9\textwidth]{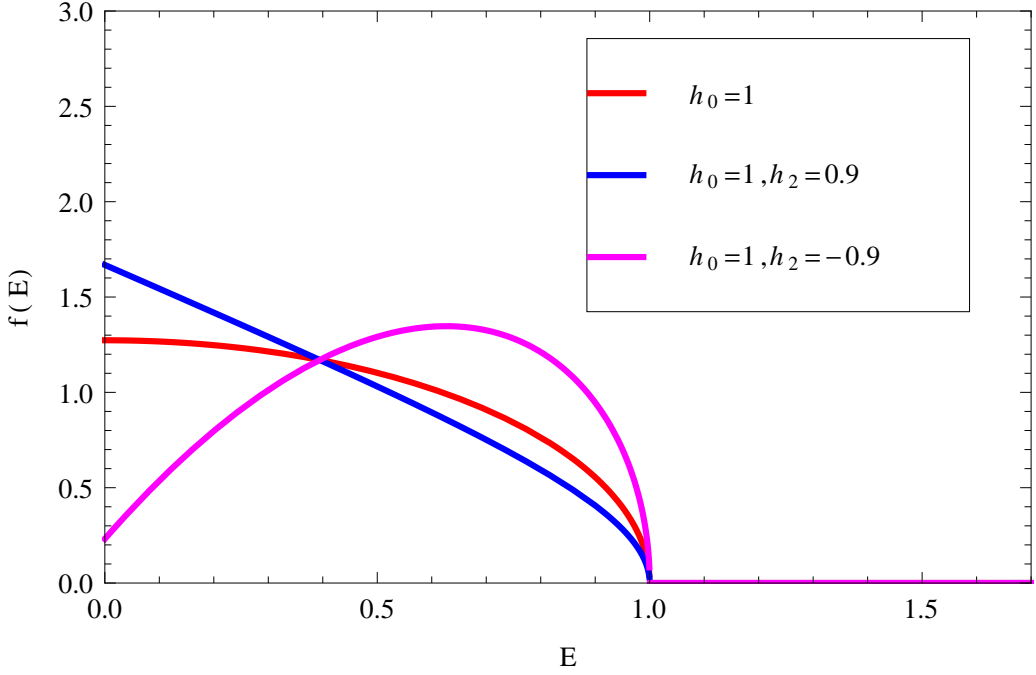}
\caption{ Energy distribution function $f[E]$ is plotted, where $E$ 
is expressed in units of $E_{m}$. The angular momentum of one of the particles
is fixed while the angular momentum of the other particle follows uniform distribution over the entire allowed range. 
The distribution function sharply declines and goes to zero at $E=E_{m}$.
The plots are made for three different cases: A. $h_0=1$,~$h_2=0$, 
B. $h_{0}=1$,~$h_2=0.9$, and C. $h_0=1$,~$h_2=-0.9$.
}
\label{dis2}
\end{center}
\end{figure}

Relaxing the assumption that angular momenta of both the particles are fixed, we now assume that 
the angular momentum of particle $1$ is $L_1^{*}$, a fixed value in the allowed range, 
while the angular momentum $L_2$ follows uniform distribution over the entire allowed range. 
\begin{equation}
g \left(L_1 \right)=\delta \left(L_1-L_1^{*}\right) \ ; \ g \left(L_2 \right)= C
 \label{h12du}
\end{equation}
Integrating over $L_1$ using Dirac's delta function, Eq.~\eq{spectrum1} can be written as
\begin{equation}
f[E] \propto  \int_{L_2^l}^{L_2^u} dL_2 \left(\frac{h_0}{\sqrt{y\left(2M-L_1^{*}\right)\left(2M-L_2\right)-E^2}}
 +\frac{h_2\sqrt{y\left(2M-L_1^{*}\right)\left(2M-L_2\right)-E^2} }{\left(y\left(2M-L_1^{*}\right)\left(2M-L_2\right)\right)}\right) \ ,
 \label{sp12du}
\end{equation}
where $L_2^l$ is the lowest value of the angular momentum stated in Eq.~\eq{L2} and $L_2^u$ is given by 
\begin{equation}
 L_2^u=2M-\frac{E^2}{y \left(2M-L_1^{*}\right)} \ .
 \label{l2u}
\end{equation}
Integrating over $L_2$, the distribution function Eq.~\eq{sp12du} can be written as 
\begin{equation}
 f[E] \propto \left(h_0+h_2\right)\sqrt{y\left(2M-L_1^{*}\right)\left(2M-L_2^l\right)-E^2} 
 -h_2 E \arctan \frac{\sqrt{y\left(2M-L_1^{*}\right)\left(2M-L_2^l\right)-E^2}}{E} \ .
 \label{sp12du1}
\end{equation}
It is clear from the expression above that the highest allowed value of the energy $E$ is 
\begin{equation}
 E_{m}=\sqrt{y\left(2M-L_1^{*}\right)\left(2M-L_2^l\right)} \ .
 \label{em2}
\end{equation}
The energy spectrum admits an upper bound $E_m$, which can go to infinity in the near-extremal limit.
The normalized distribution function can be written as
\begin{equation}
 f[E]=\frac{4}{\pi \left(h_0+\frac{h_2}{2}\right)}\frac{1}{E_{m}^2}\left( \left(h_0+h_2\right)\sqrt{E_m^2-E^2}-h_2 E\arctan\frac{\sqrt{E_m^2-E^2}}{E}\right) \ .
 \label{nd2}
\end{equation}
The energy distribution function goes to zero very sharply when the energy approaches highest energy $E=E_{m}$. 
The divergence of the distribution when both the angular momenta were fixed disappears even when one of them is allowed 
to vary over different values.
The behavior is qualitatively different for positive and negative values of the parameter $h_2$ which depicts the anisotropic
emission of the massless particles in the center-of-mass frame.

\subsection{Both $L_1$ and $L_2$ follow uniform distribution}

\begin{figure}
\begin{center}
\includegraphics[width=0.9\textwidth]{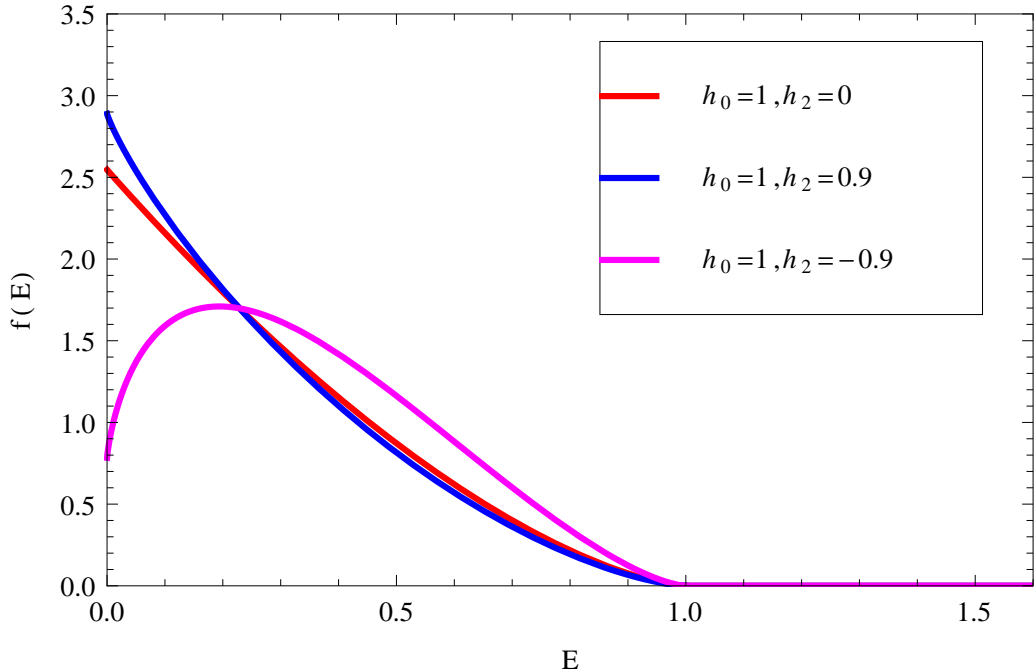}
\caption{ Energy distribution function $f[E]$ is plotted here, where $E$  
is expressed in units of $E_{m}$. The angular momenta of both the particles
follow uniform distribution over the entire allowed range, which is
 a physically realistic assumption. 
The distribution function goes to zero at $E=E_{m}$ rather slowly and smoothly.
Plots are made for three different cases: A. $h_0=1$,~$h_2=0$,
 B. $h_{0}=1$ ,~$h_2=0.9$, and C. $h_0=1$,~$h_2=-0.9$.
The distribution is drastically different, depending on whether $h_2$ is positive or negative.
}
\label{dis3}
\end{center}
\end{figure}

We now make the assumption that the angular momenta of both the particles are uniformly distributed over the entire allowed range. 
This is a quite realistic assumption. That is,
\begin{equation}
g \left(L_1 \right)= C_1 \ \text{and} \ g \left(L_2 \right)= C_2 \ .
 \label{h12uu}
\end{equation}
In this case both the integrals over $L_1$ and $L_2$ are non-trivial.
\begin{equation}
 f\left[E\right] \propto \int_{L_1^l}^{L_1^u} dL_1 \int_{L_2^l}^{L_2^u(L_1)} dL_2
 \left(\frac{h_0}{\sqrt{y\left(2M-L_1\right)\left(2M-L_2\right)-E^2}}+\frac{h_2\sqrt{y\left(2M-L_1\right)\left(2M-L_2\right)-E^2} }{\left(y\left(2M-L_1\right)\left(2M-L_2\right)\right)}\right)
\label{sp12uu}
 \end{equation}
The lower limits of integration $L_1^l$ and $L_2^l$ are the lowest permissible values of the angular momenta discussed in 
Eqs.~\eq{L1} and \eq{L2}. While $L_1^u$ and $L_2^u(L_1)$ are given by 
\begin{equation}
 L_1^u=2M-\frac{E^2}{y \left(2M-L_2^l\right)} \ \text{and} \  L_2^u(L_1)=2M-\frac{E^2}{y \left(2M-L_1\right)} \ ,
 \label{l12l}
\end{equation}
respectively. Integrating over $L_1$ and $L_2$, we can write the distribution function \eq{sp12uu} as
\begin{eqnarray}
 f[E] \propto  && 2\left(h_0+h_2\right) \left( \sqrt{y\left(2M-L_1^l\right)\left(2M-L_2^l\right)-E^2} 
 -E \arctan \frac{\sqrt{y\left(2M-L_1^l\right)\left(2M-L_2^l\right)-E^2}}{E} \right) \nonumber \\
 && - i h_2 E  \left(\frac{\pi^2}{24}+ \frac{1}{2}\left(\arctan \frac{\sqrt{y\left(2M-L_1^l\right)\left(2M-L_2^l\right)-E^2}}{E}\right)^2\right) \nonumber \\
 && +h_2 E \ \arctan \frac{\sqrt{y\left(2M-L_1^l\right)\left(2M-L_2^l\right)-E^2}}{E} \log \left(1+ e^{2 i \arctan \frac{\sqrt{y\left(2M-L_1^l\right)\left(2M-L_2^l\right)-E^2}}{E}}\right) \nonumber \\
 && -i h_2 E \ \frac{1}{2} \text{PolyLog}  \left(2,-e^{2 i \arctan \frac{\sqrt{y\left(2M-L_1^l\right)\left(2M-L_2^l\right)-E^2}}{E}}\right) \ .
 \label{sp12uu2}
\end{eqnarray}
The upper limit on the allowed value of the energy $E$ is 
\begin{equation}
 E_{m}=\sqrt{y\left(2M-L_1^l\right)\left(2M-L_2^l\right)} \ .
 \label{em3}
\end{equation}
Thus, we obtain an upper cut-off $E_m$, which can go to infinity in the near-extremal limit.

The normalized distribution can be written as
\begin{equation}
 f(E)=\frac{4}{\pi \left(h_0+\frac{3}{4}h_2\right)}\frac{1}{E_{m}^2}\left( 2\left(h_0+h_2\right)\left(\sqrt{E_m^2-E^2}- E\arctan\frac{\sqrt{E_m^2-E^2}}{E}\right)-h_2 E \  \Sigma(E)     \right) \ ,
 \label{nd3}
\end{equation}
where
\begin{eqnarray}
\Sigma(E)= && i\frac{\pi^2}{24}+ i \frac{1}{2}\left(\arctan\frac{\sqrt{E_m^2-E^2}}{E}\right)^2 - \arctan\frac{\sqrt{E_m^2-E^2}}{E} \log\left(1+e^{2i\arctan\frac{\sqrt{E_m^2-E^2}}{E}}\right) \nonumber \\
&& + i \frac{1}{2} \text{PolyLog} \left(2,-e^{2i\arctan\frac{\sqrt{E_m^2-E^2}}{E}}\right) \ .
 \label{nd3def}
\end{eqnarray}

The distribution function goes to zero rather slowly and smoothly as the highest energy $E=E_m$ is reached as compared to the previous
case, where one of the angular momentum followed Dirac's delta distribution. The distribution is concave for non-negative values of 
$h_2$. For negative values of $h_2$ it becomes concave as energy approaches $E=E_{m}$. The distribution was always convex when one of the angular
momentum was fixed. The behavior is quite different in the two cases, where $h_2$ is positive and negative. This implies that the 
anisotropy in the emission of the massless particles which is dictated by the differential cross section of the underlying 
particle physics process is imprinted on the spectrum in a subtle
way. Thus, the observation of the spectrum of ultra-high-energy particles can allow us distinguish different particle physics models at ultra-high energies where particles collide.

In all three cases, the energy distribution function depends only on $E_m$, $h_0$ and $h_1$. As mentioned earlier $h_0$ and $h_1$
are dictated by the fundamental physics. $E_m$ which is the upper bound on the energy of the massless particle, is dictated by 
the mass of the colliding particle $m$ and the parameter $\epsilon$ depicting the deviation of the Kerr spin parameter from
extremal value. From \eq{em1}, \eq{em2} and \eq{em3}, $E_m$ turns of to be $O(\frac{m}{\sqrt{\epsilon}})$. We rewrite it as 
\begin{equation}
\frac{m}{\sqrt{\epsilon}}=\left(\frac{m}{m_p}\right) \frac{1}{\sqrt{\epsilon}} \text{GeV}
 \label{emap}
\end{equation}
where $m_p$ is the mass of proton. For $E_m$ to be comparable to the astrophysically relevant values $\epsilon$ must be very small.
For $E_m$ to correspond to the maximum energy of the ultra-high energy cosmic rays $10^11 \text{Gev}$ and that of ultra-high energy 
neutrinos $10^{6} \text{GeV}$, we must have $\epsilon=10^{-22}$ and $\epsilon=10^{-12}$ respectively, if we assume that the 
massless particles are produced in the collision of proton-like particles. As we explain in the next section, the over-spinning 
Kerr geometry is driven towards the extremality as a consequence of the collisional Penrose process. Thus such small
values of the parameter $\epsilon$ can be realized quite naturally.

\section{Upper bound on the energy}

The expressions for the conserved energy and angular momentum of the massless particle produced in the collision, from \eq{EU}
and \eq{Lp}, are given by 
\begin{equation}
 E=\frac{m}{\sqrt{2}M} \frac{\sqrt{2M-L_2}\sqrt{2M-L_1}}{\sqrt{\epsilon}} \sin \alpha \ ,
 \label{EU1}
\end{equation}
and
\begin{equation}
 L=\sqrt{2}m \frac{\sqrt{2M-L_2}\sqrt{2M-L_1}}{\sqrt{\epsilon}} \sin \alpha \ ,
 \label{Lp2}
\end{equation}
respectively.
If the massless particle is emitted in the upper plane in the center of mass frame away from $\hat{r}$ axis, 
i.e. when $\alpha \in (0,\pi)$, its conserved energy as well as conserved angular momentum is extremely large. 
Whereas for the massless particle which is emitted in the diametrically opposite direction with respect to the first 
particle in the lower plane for which $\alpha \in (\pi,2\pi)$, both conserved energy as well as the angular momentum
takes a value which is large and negative. 

Particle emitted in the upper plane away from $\hat{r}$ axis escapes and carries large energy and angular momentum to infinity. 
Whereas the particle emitted in the lower plane with large negative energy and angular momentum eventually hits the singularity,
thereby reducing the Kerr mass and angular momentum parameters. 

The conserved energy $E$ and angular momentum $L$ of the massless particle escaping to infinity are related by 
\begin{equation}
 L=2M E \ .
 \label{bb1}
\end{equation}
Thus final Kerr mass parameter $M_{f}$ and spin parameter $a_{f}$ are given by 
\begin{equation}
 M_{f}=M-E ~~~ \text{and} ~~~ a_{f}=\frac{Ma-L}{M-E} \ .
 \label{amf}
\end{equation}
It is clear from the expression above that the final spin parameter is smaller than its initial value, i.e., $a_{f}<a$.
Hence the collisional Penorse process leads to the reduction of the Kerr spin parameter and the over-sppinning Kerr 
geometry is driven towards the extremality when the back-reaction is taken into account. 
Thus the rotational energy is being extracted from the over-spinning Kerr spacetime as in the case of Kerr black hole. 

The present calculation is meaningful so along as the final configuration is over-spinning Kerr geometry. The final dimensionless
spin parameter can be obtained from \eq{amf} and is given by 
\begin{equation}
\frac{a_{f}}{M_{f}}=\frac{1-\frac{2E}{M}+\epsilon}{1-\frac{2E}{M}+\frac{E^2}{M^2}} \ ,
\end{equation}
where we have used the relation $a=M(a+\epsilon)$. It must be larger than unity. Thus we get 
\begin{equation}
 \epsilon > \frac{E^2}{M^2} \ .
 \label{dimf}
\end{equation}
Eliminating $\epsilon$ from \eq{EU1} and \eq{dimf}, we get an upper bound on the energy of the massless particle as 
\begin{eqnarray}
 E &<& \left(1-\frac{L_1}{2M}\right)^{\frac{1}{4}} \left(1-\frac{L_2}{2M}\right)^{\frac{1}{4}}\sqrt{2 \sin \alpha} \times \sqrt{mM} \ .
\end{eqnarray}
Thus the energy of the massless particle produced is at most $O(\sqrt{mM})$. We rewrite $\sqrt{mM}$ as 
\begin{equation}
 \sqrt{mM}=3.2 \times 10^{28} \left(\frac{m}{m_{p}}\right)^{\frac{1}{2}} \left(\frac{M}{M_{\odot}}\right)^{\frac{1}{2}} \text{GeV} \ ,
\end{equation}
where $m_p$ and $M_{\odot}$ are the masses of proton and sun respectively. As we stated earlier, the energy of ultra-high energy cosmic rays 
is upto $10^{11}$ GeV and that of the ultra-high energy neutrinos is upto $10^{6}$ GeV. Thus the upper bound is still so high that it can be of great
interest from astrophysical and particle physics point of view.

\section{Conclusion}
In this paper we presented a novel mechanism to generate the ultra-high-energy particles starting with particles with the moderate energies
that is exclusively based on gravity exploiting the collisional Penorse process. All other acceleration mechanisms proposed so far make use of electromagnetic interaction to 
accelerate particles to high energy. We consider an overspinning Kerr geometry transcending the extremality by an infinitesimal amount 
$a=M(1+\epsilon)$, where we take the limit 
$\epsilon \rightarrow 0^+$. 
Overspinning spacetime geometries occur quite naturally in the context of string theory and also may appear as a transient 
configuration in the process of gravitational collapse of regular matter cloud. We showed that the ultra-high-energy collisions are necessary to produce 
the particles with large energies in the Kerr spacetime. Collisions with divergent center-of-mass energy can occur around the near-extremal Kerr
black holes as well as in the overspinning Kerr spacetime geometry. In this paper we demonstrate that the efficiency of the collisional Penorse
process can diverge in the overspinning Kerr geometry in near-extremal
limit $\eta \sim  {1}/{\sqrt{\epsilon}} \rightarrow \infty$, while the efficiency 
is always finite in the context of the Kerr black holes.

We consider two identical massive particles that start from rest at infinity and fall towards the singularity. 
One of the particles turns back at $r<M$ as it encounters the angular momentum barrier and appears at $r=M$ as an outgoing particle. 
The other particle appears at $r=M$ as an ingoing particle where it collides with an outgoing particle. The center-of-mass energy of collision 
shows divergence in the near-extremal limit. We considered a process where the two colliding particles are scattered to produce  
two massless particles. All the particles were assumed to move on the equatorial plane, for simplicity, since it allows to carry out 
a fully analytical analysis of the collisional Penrose process. We made a transition to the center-of-mass tetrad since 
the differential cross section of the underlying particle physics process is expressed in the center-of-mass frame, which makes 
the calculation of the spectrum feasible. By analyzing the null geodesics, we identify the escape cones, i.e., the set of directions 
in the center-of-mass frame along which the massless particle must be emitted so that it escapes to infinity. We show that the 
escape cones span almost half of the entire angular range.
We compute the conserved energy of the massless particle, which is its energy as measured by an asymptotic observer if it escapes to infinity. 
We show that particles that escape along almost all the angles within the escape cones have divergent energies, while the particles 
that escape with finite energies are in minuscule minority. The conserved energy of the particles emitted in the opposite direction 
to that of the ultra-high-energy particles reaching infinity tends to negative infinity. This implies that the collisional Penrose process is at work 
and is responsible for the generation of ultra-high-energy particles in the overspinning Kerr geometry. 
These results do not depend on whether the center-of-mass frame moves radially outwards, radially inwards or 
does not admit any motion in the radial direction. 

We show that all the ultra-high-energy particles have positive angular momenta and almost the same values of the impact parameter and 
thus the particles co-rotate with the singularity. A distant observer sees the ultra-high-energy particles originating from a 
bright spot which is located at the specific location on that side of the singularity which is rotating towards the observer. We computed the 
spectrum of the ultra-high-energy particles assuming that the cross section of the 
underlying particle physics process is constant at the large center-of-mass energies
and distribution of the angular momenta of the colliding particles is uniform.
We consider the case where the emission of the massless particles in the center-of-mass frame is isotropic as well the case 
where the emission is anisotropic with the anisotropy parametrized in a specific way. We find that there is an upper bound
on the energy of the massless particle. The upper bound however can go to infinity in the near-extremal limit. The functional 
form of the energy distribution function is different above and below the upper bound.
We show that the anisotropy of emission in the center-of-mass frame has its signature imprinted on the spectrum. Since 
the anisotropy is dictated by the differential cross section of the particle physics process, the observation of the spectrum 
can put constraints on and distinguish different particle physics models at high energies. 

Thus ,~a near-extremal overspinning Kerr geometry will allow us to kill two birds with one stone. Firstly it would provide a mechanism 
to generate ultra-high-energy particles starting from the particles with moderate energies employing the collisional Penrose process
with divergent efficiency and secondly the observation of the spectrum of high-energy particles would allow us to distinguish
the different particle physics models. Hence, the existence of near-extremal overspinning Kerr geometry either as a permanent or 
a transient configuration would have a deep impact on the astrophysics as well as the fundamental particle physics.

If the backreaction is taken into account then the over-spinning Kerr geometry is driven towards the extremal Kerr black hole
configuration. The calculation in this paper is meaningful only so long as the final spin parameter is larger than the extremal 
value. This puts an upper bound on the energy of the massless particle. We show that the upper bound is still so high that it can be of great
interest from astrophysical and particle physics point of view.

\section*{acknowledgement}
T.H. and M.P. would like to thank T.~Igata and K.~Ogasawara 
for fruitful discussion. 
M.K. is partially supported by a grant for research abroad from JSPS.
M.K. acknowledges financial support provided under the European
Union's H2020 ERC Consolidator Grant ``Matter and strong-field gravity: New frontiers in Einstein's theory''
grant agreement no. MaGRaTh-646597, and under the H2020-MSCA-RISE-2015 Grant No. StronGrHEP-690904.
M.P. was supported by the Research Center for Measurement in Advanced Science in Rikkyo University.
This work was partially supported by JSPS KAKENHI Grant Number 26400282.

\end{document}